\documentclass[a4paper,12pt]{article}
\usepackage{stolenstyle}
\usepackage{amsmath,epsf,amssymb,latexsym,amsthm,setspace,array,pifont,color,hyperref,amsfonts,dsfont,cancel,braket,parskip,verbatim,slashed, hyperref,graphicx,feynmp-auto}
\usepackage[usenames,dvipsnames]{xcolor}
\usepackage[framemethod=default]{mdframed}
\usepackage{avant} 
\hypersetup{
	colorlinks,
	urlcolor=Bittersweet,
	linkcolor=Bittersweet,
	citecolor=Bittersweet
}
\usepackage[none]{hyphenat} 
\graphicspath{{Pictures/}} 

\newcommand{\db}{\dot{b}}

\newcommand{\R}{\mathbb{R}}
\newcommand{\C}{\mathbb{C}}

\newcommand{\cl}[1]{\mathcal{#1}}

\newcommand{\Trm}[1]{\text{Tr$_-$}\left(#1\right)}

\newcommand{\Trpm}[1]{\text{Tr$_\pm$}\left(#1\right)}
\newmdenv[skipabove=7pt,
skipbelow=7pt,
rightline=false,
leftline=false,
topline=false,
bottomline=false,
backgroundcolor=gray!10,
linecolor=gray,
innerleftmargin=5pt,
innerrightmargin=5pt,
innertopmargin=5pt,
innerbottommargin=5pt,
leftmargin=0cm,
rightmargin=0cm,
linewidth=4pt]{eBox}

\def\spvec#1{\left(\vcenter{\halign{\hfil$##$\hfil\cr \spvecA#1;;}}\right)}
\def\spvecA#1;{\if;#1;\else #1\cr \expandafter \spvecA \fi}

\makeatletter
\renewcommand*\env@matrix[1][*\c@MaxMatrixCols c]{%
	\hskip -\arraycolsep
	\let\@ifnextchar\new@ifnextchar
	\array{#1}}
\makeatother
\def\preprint{ {{~}}}
\title{Amplitudes for Astrophysicists: Known Knowns}
\author[1,2] {Daniel J. Burger,}
\author[1,2,3,4]{Ra{\'u}l Carballo-Rubio,}
\author[1,2] {Nathan Moynihan,}
\author[1,5] {Jeff Murugan}
\author[2,5,6] {and Amanda Weltman}
\affiliation[1]{The Laboratory for Quantum Gravity \& Strings,}
\affiliation[2]{The Cosmology \& Gravity Group,\\ Department of Mathematics and Applied Mathematics,\\
	University of Cape Town,\\
	Private Bag, Rondebosch, 7700,\\
	South Africa.}
\affiliation[3]{SISSA, International School for Advanced Studies,\\ Via Bonomea 265, 34136 Trieste, Italy}
\affiliation[4]{INFN Sezione di Trieste, Via Valerio 2, 34127 Trieste, Italy}
\affiliation[5]{School of Natural Sciences,\\
	Institute for Advanced Study, Olden Lane, Princeton,\\
	NJ 08540, USA}
\affiliation[6]{Department of Astrophysical Sciences, \\ Peyton Hall, Princeton University, Princeton, \\NJ  08544, USA}

\emailAdd{burgerj.daan@gmail.com}
\emailAdd{raul.carballorubio@sissa.it}
\emailAdd{nathantmoynihan@gmail.com}
\emailAdd{jeff.murugan@uct.ac.za}
\emailAdd{amanda.weltman@uct.ac.za}
\abstract{
The use of quantum field theory to understand astrophysical phenomena is not new. However, for the most part, the methods used are those that have been developed decades ago. The intervening years have seen some remarkable developments in computational quantum field theoretic tools. In particle physics, this technology has facilitated calculations that, even ten years ago would have seemed laughably difficult. It is remarkable, then, that most of these new techniques have remained firmly within the domain of high energy physics. We would like to change this. As alluded to in the title, this paper is aimed at showcasing the use of modern on-shell methods in the context of astrophysics and cosmology. In this article, we use the old problem of the bending of light by a compact object as an anchor to pedagogically develop these new computational tools. Once developed, we then illustrate their power and utility with an application to the scattering of gravitational waves.
}
\begin{document}
	\maketitle
	\date{today}
\begin{fmffile}{amplitudes}

\arraycolsep=10pt\def\arraystretch{2.2}

\newpage
\section{Introduction}

There are moments in the evolution of scientific research when the infusion of technology from adjacent research fields profoundly changes the game. In physics for example, the linear algebra of matrices in quantum mechanics, Riemannian differential geometry in general relativity, Feynman diagrams in condensed matter and, more recently, Maldacena's use of conformal field theory methods to compute cosmological non-Gaussianities \cite{Maldacena:2002vr,Maldacena:2011nz} are among those that immediately come to mind. In each case, well-developed techniques from one field, brought to bear on select problems of another were used to leapfrog the latter well beyond what was thought possible at the time. It is our hope that we are on the cusp of another such development today.

Indeed, the last decade has seen a somewhat silent revolution play out in quantum field theory that, at the very least, has changed the way basic quantities like scattering amplitudes are calculated but, more likely than not, will change (yet again) the way we think about space, time, locality and causality. For the want of a better term, we will call this the {\it amplitudes revolution} and, while its sparks were cast in the 90's already, it was not until a remarkable insight by Witten in 2004 \cite{Witten:2003nn}, showing how to compute scattering amplitudes using a new (to most string theorists anyway) set of twistor variables, that saw the field ignite. Almost overnight, it was realized that use of these variables could drastically simplify the computation of scattering problems by effectively reducing the evaluation of integrals encoded in Feynman diagrams (themselves having ``brought computation to the masses" back in the 50's when they ushered in the era of particle physics) to recursion relations and consequently whittling down hundreds of pages of calculation down to a mere handful (see, for example, page 484 of \cite{Zee:2003mt}). 
The implications of this simplification are profound. In {\it particle phenomenology}, for example, where typical disambiguation of background events from wanted scattering events (the recent discovery of the Higgs at the LHC should spring to mind) requires the evaluation of literally hundreds of diagrams on thousands of pages, the amplitude revolution was a gift of messianic proportions\footnote{In the sense that sometime in the near, but unspecified, future things are going to get {\it really} good.}. On the other side of town, in more formal {\it quantum field theory}, amplitude methods have introduced fantastic new word combinations like ``correlators to ten loops" \cite{Bourjaily:2015bpz,Bourjaily:2016evz} to the phrasebook of theorists that only a few years ago would have been as ridiculed as other recent lexiconic anomalies like ``muggle", ``hackathon" and ``cyberslacking". Today, they represent a new frontier of quantum field theory. Nor have these methods gone unnoticed by the {\it gravity} community where they have been used to demonstrate the remarkable fact that general relativity behaves like the square of Yang-Mills gauge theory \cite{Kawai:1985xq,Bern:2002kj,Bern:2008qj}. Nevertheless, you might wonder, what does all of this have to do with {\it astrophysics}, a decidedly practical research field rooted in observations in the sky? 
 
To answer this, we take the view that, like particle physics, astrophysics is partially about scattering events. However, while particle physics is, by definition, concerned with the subatomic world, astrophysics involves the study of the scattering of light off decidedly macroscopic objects. Fortunately, the scattering formalism cares little for this distinction and many landmark astrophysical problems may be treated very effectively as photon-scattering processes, their amplitudes computed and a remarkable amount of physical information extracted. Moreover, as a formalism purpose built to deal with quantum problems perturbatively, scattering amplitudes come with a systematic way to deal with quantum corrections which more traditional methods are simply unable to do. Beyond scattering of light, the recent confirmation of gravitational waves \cite{Abbott:2016blz} heralds a new era of gravitational wave astronomy and presumably also astrophysics \cite{Buonanno:2014aza}. Here again, the power and versatility of the amplitude formalism shines through. Apart from some relatively minor technical modifications, all that is required is the trading in of one massless probe particle, the photon, with another, the graviton.

In this article, we introduce the reader (who we will assume is an astrophysicist) to the main tools of the trade, the spinor-helicity formalism and the BCFW recursion relations\footnote{Named after its discoverers, Britto-Cachazo-Feng and Witten.}. In the interests of pedagogy, we do so with one of the benchmark problems in astrophysics, the gravitational lensing of light by a massive celestial body. Indeed, the example of gravitational lensing forms such an integral part of this paper that we might well have titled it ``Gravitational lensing, three ways". Not only is the answer to this problem known\footnote{This explains, incidentally,  the reference to the Rumsfeld classification of knowledge in the title.} \cite{Einstein:1956zz}, but we are not even the first to frame the problem in terms of photon scattering \cite{Bjerrum-Bohr:2014zsa,Bjerrum-Bohr:2016hpa,Bjerrum-Bohr:2017dxw,Guevara:2017csg}. Other relevant works in the literature include the construction of radiating solutions by means of the classical double copy \cite{Goldberger:2016iau,Goldberger:2017frp} and of classical spacetimes from amplitudes \cite{Luna:2017dtq,Neill:2013wsa}, or also the application of quantum field theory techniques to gravitational lensing \cite{Jenkins:2014hza}. We are however, to the best of our knowledge, the first to embrace this new technology as a broadly applicable tool in astrophysics. To illustrate this, in section \ref{sec:gravscal} we use the BCFW relations to study the problem of gravitational wave lensing. 

This article is our attempt to expose an astrophysical audience with a modest background in quantum mechanics and general relativity (and a dash of quantum field theory would not hurt) to some of these powerful tools of contemporary field theory. Where at all possible, we will make available the necessary mathematical and physical ideas necessary to follow a particular argument in the main text. However, a detailed presentation of all the multitude nuances of these techniques is beyond the scope of this article. To compensate, we have appended to the main text a series of useful appendices that contain some relevant mathematics, physics, a glossary of terminology and a detailed list of  references that we ourselves found useful. We would like to emphasize that this article is not a monograph on the topic, nor is it meant to be a comprehensive account of the latest developments, or even of the literature, in the field. It should be regarded simply as an invitation to a different way of thinking about problems, some old, some new, in astrophysics.

\subsection{Notation and conventions}

In what follows, we will use a ``mostly plus" 4-dimensional Minkowski metric $(-,+,+,+)$  
and, unless otherwise stated, work in natural units where $\hbar = c = 1$. A particle is said to be on-shell if its momentum satisfies the condition $p_\mu p^\mu \equiv p^2 = -m^2$.

The interaction between $n$ external particles is described by an $n$-point amplitude and denoted 
by $A(1^{h_1}2^{h_2}\cdots n^{h_n})$. Such amplitudes are typically labelled by the momenta $p^{\mu}_{i}$ and helicity $h_{i}$ of the external particles. For example, $2^{h_{2}}$ in the preceding amplitude refers to external particle 2 which carries momentum $p^{\mu}_{2}$ with corresponding helicity $h_{2}$. By convention, we will also assume that all momenta are outgoing so that momentum conservation of $n$ external particles
implies that
\begin{equation}
  \sum_i^n p_i = 0.\label{eq:momcons}
\end{equation}

Finally, it will sometimes be more convenient to use other variables instead of the momenta themselves in order to write $n$-point amplitudes. One set that will feature prominently, are the so-called Mandelstam invariants
 \begin{equation}
 s_{ij} = -(p_i + p_j)^2.\label{eq:Mandelstam}
 \end{equation}

\section{Light bending}
Having dispensed then with the formalities, let's set the scene by thinking about a century old problem that is probably to general relativity what the double slit experiment is to quantum mechanics; the gravitational lensing of light by a massive body.

\subsection{Gravitational lensing as an exercise in general relativity \label{sec:glgr}}
We will start by briefly reviewing the classical deflection of light due to a massive body in the context of General Relativity. 
To determine the path of a massless particle, the photon in this case, in a gravitational field, we need to know the {\it null geodesics} in the corresponding spacetime. We reasonably assume a stationary spacetime with spherical symmetry which naturally leads us to the Schwarzschild geometry. The metric of the Schwarzschild solution for a body of mass M is
\begin{equation}
ds^2 = -\left(1-\frac{2M}{r}\right)dt^2 +  \left(1-\frac{2M}{r}\right)^{-1}dr^2 +r^2( d\theta^2 + \sin^2\theta\,d\varphi^2). \label{Schwarzschild}
\end{equation}
First it is necessary to recall that if we define the the tangent to a geodesic as $u^{\alpha}=dx^{\alpha}/d\lambda$ where $\lambda$ is an affine parameter along the geodesic, the inner product of $u^\alpha$ with the Killing field of the geometry, $u^{\mu}\xi_{\mu}$ is a constant. 
In fact, from this we can read off the constants of motion,
\begin{align} \label{GR_Constants of Motion}
u\cdot\xi = u^t \xi_t +  u^\varphi \xi_\varphi = \left(1-\frac{2M}{r}\right)\frac{dt}{d\lambda} + (r^2 \sin^2\theta)\frac{d\varphi}{d\lambda} = E +L,
\end{align}
since, if the affine parameter, $\lambda$, is normalized such that $u^{\alpha}$ coincides with the momentum of a null vector then, $E$ and $L$ are to be understood as the energy and the angular momentum of the photon respectively. The rotational symmetry of the Schwarzschild metric implies that if a null geodesic starts in, say, the equatorial plane then the entire geodesic remains in the plane, meaning we can set $\theta = \pi/2$ without loss of generality. We require that the tangent vector be null which, by the geodesic equation, tells us that
\begin{align}
0 = g_{\mu\nu}\frac{dx^{\mu}}{d\lambda}\frac{dx^{\nu}}{d\lambda} =
-\left(1-\frac{2M}{r}\right)\left(\frac{dt}{d\lambda}\right)^2 + \left(1-\frac{2M}{r}\right)^{-1}\left(\frac{dr}{d\lambda}\right)^2 + r^2 \left(\frac{d\varphi}{d\lambda}\right)^2\,,
\end{align}
which after plugging in Eq. \eqref{GR_Constants of Motion} and with a little algebraic manipulation gives,
\begin{align} \label{GR_Energy Relation}
\frac{1}{2}E^2 = \frac{1}{2}\left(\frac{dr}{d\lambda}\right)^2 + \frac{L^2}{2 r^2 }\left(1-\frac{2M}{r}\right).
\end{align}

The deflection angle of a light ray is usually framed in terms of the {\it impact parameter} which, in flat space is defined by $b=L/E$. Since we consider only paths in the weak field regime, $r\gg M$, $b$ will serve as our apparent impact parameter. From eq. \eqref{GR_Energy Relation}, we find that the effective potential of massless orbits is $V(r) = L^2(r-2M)/2 r^3$. We next define the point of closest approach of the particle to the center of the geometry as $r=R_0$, the point at which the photon will have a turning point as it passes near the massive body.

From equation \eqref{GR_Energy Relation} and using the definition of the angular momentum that, for $\theta=\pi/2$, takes the form $L=r^2d\varphi/d\lambda$ we find 
\begin{equation} \label{GR_DPDR}
\frac{d\varphi}{dr} = \frac{L}{r^2}\left( E-\frac{L^2}{r^3}(r-2M)\right)^{-1/2}.
\end{equation}
This is now sufficient for us to calculate the change in the photon's trajectory. Assuming that the particle approaches from and proceeds to infinity,  this change is captured by the angle between these trajectories as a result of the deflection of a photon due to gravity, $\Delta\varphi = \varphi_{+\infty} - \varphi_{-\infty}$. However, as a consequence of the symmetries of the geometry, the contributions to the integrals before and after the turning point are equal. To determine the opening angle $\Delta\varphi$, we have to integrate eq. \eqref{GR_DPDR}. this is easiest to do by introducing the new variable $u=1/r$ and using the effective potential to eliminate $b$ to give
\begin{equation}
\Delta\varphi = 2\int^{1/R_0}_{0} du \left( R_0^{-2}-2M R_0^{-3}-u^2+2M u^3  \right)^{-1/2}.
\end{equation}
As a check, we set $M=0$ to give
\begin{equation}
\Delta\varphi = 2\arcsin 1 =\pi,
\end{equation}
which is of course the expected result in flat spacetime. Evaluating the integral for $M\neq0$ to first order in $M$, we need to treat $M$ and $R_0$ as independent variables, and then vary the integrand with respect to $M$. This allows us to calculate the deflection angle to first order in $M$ as a function of mass but at a fixed radius $R_0$. It is important to note that the physical parameter we want in the result is the impact parameter, but if $M=0$ we have $b=R_0$. Differentiating with respect to $M$ and evaluating the result at $M=0$ gives
\begin{equation}
\left.\frac{\partial \varphi}{\partial M}\right|_{M=0}=
2\int_{0}^{1/b} du \left( b^{-3}-u^3 \right) \left(b^{-2}-u^2 \right)^{-3/2}=\frac{4}{b}.
\end{equation}
So, to first order in $M$,
\begin{equation}
\Delta\varphi= \pi + M\left.\frac{\partial \varphi}{\partial M}\right|_{M=0}=\pi + \frac{4M}{b}.
\end{equation}
Of course, in order to compute the deflection angle, we are interested in the deviation from the flat spacetime trajectory induced by the Schwarzschild geometry, {\it i.e.}
\begin{equation}
\varphi_{\rm D} = \Delta\varphi -\pi = \frac{4GM}{R_{0}}.\label{eq:scattangle}
\end{equation}
\begin{figure}[h]
\begin{center}
	\includegraphics[width=0.85\textwidth]{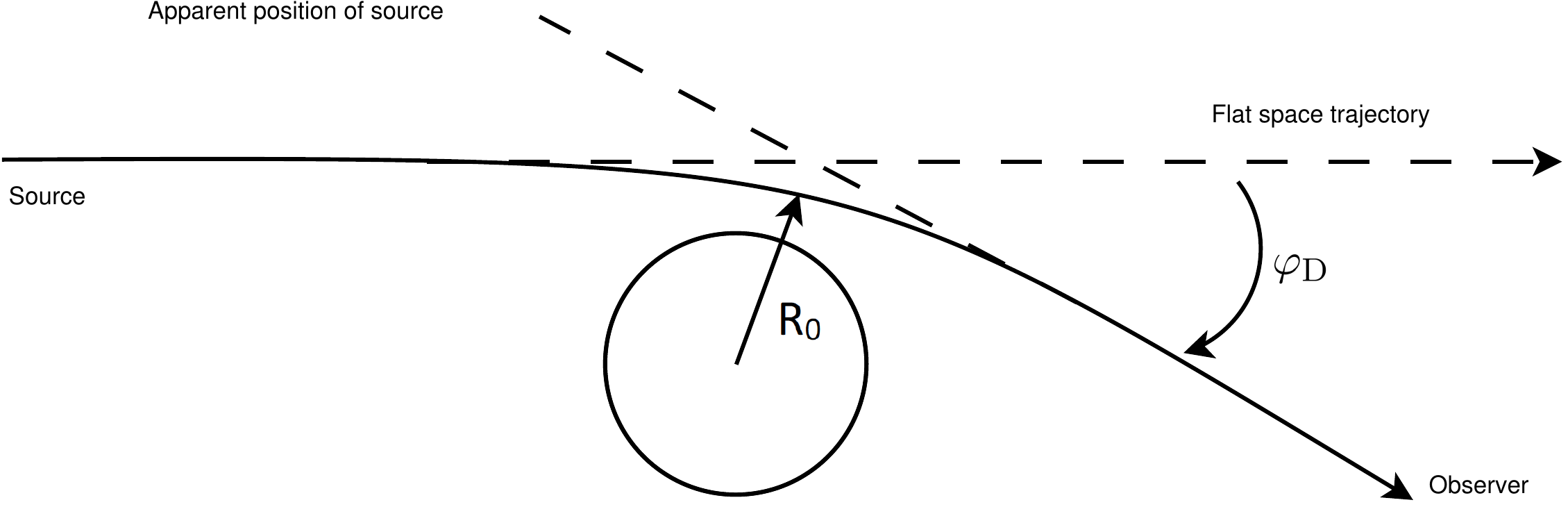}
	\vspace{0.5cm}
	\caption{Graphical representation of the deflection angle $\varphi_{\rm D}$.}
\end{center}
\end{figure}

At this point it is worth noting that apart from assuming that the lensed beam of photons obeys the weak field Einstein equations and moves on a null geodesic in the geometric optics limit, we did not have to specify anything else about the probe. Essentially then, the derivation of the deflection angle is applicable to any massless particle moving in a stationary, spherically symmetric space time exterior to a massive object with mass $M$. This is an important point that we will return to shortly.

\subsection{Gravitational lensing as an exercise in quantum field theory \label{sec:lbqft}}

To illustrate the utility of quantum field theory in astrophysical phenomena, let's look back at the same physics in a slightly different light. In this subsection, we will calculate the light bending angle that corresponds to a photon interacting gravitationally with a massive object, for example a star. In the language of quantum field theory {\it classical processes} like this correspond to tree level interactions whose Feynman diagrams look like, well, trees. {\it Quantum corrections} to these results arise from   interactions internal to a particular diagram that are mediated by virtual particles. These arise as loops in the diagram and correspond to some or other (usually difficult) integration that needs to be carried out. For a concise review of how to translate between Lagrangians, Feynman diagrams and observables, we refer the reader to \cite{Kumericki:2016} for a simple introduction, or to \cite{Zee:2003mt,Peskin:1995ev,Srednicki_2007} for a more in depth treatment. The aim of this subsection is merely to show that thinking about light-bending as a tree-level interaction between photons and a massive neutral scalar particle at tree level mediated by Einstein gravity allows us to derive a light-bending angle in perfect agreement with the result obtained from general relativity.

Given the ingredients of the physical process that we're interested in, it makes sense to start with the Einstein-Maxwell-scalar action,
\begin{equation}
S[A,h,\phi] = \int d^4x \sqrt{-g}\left(\frac{2}{\kappa^2}R -\frac{1}{4} g^{\mu\nu}g^{\alpha\beta}F_{\mu\alpha}F_{\nu\beta} + g^{\mu\nu}\partial_\mu\phi\partial_\nu\phi - \frac{m^2}{2}\phi^2\right)\,.\label{eq:EMSact}
\end{equation}
In quantum field theory, this type of interaction is one where an incoming photon exchanges a virtual graviton with a scalar object. With time increasing left to right, the corresponding Feynman diagram would be schematically given by\\
\begin{equation}
\begin{fmfgraph*}(160,60)
\fmfleft{i1,i2}
\fmfright{o1,o2}
\fmf{photon,label.side=left}{i1,v1}
\fmf{plain,label.side=right}{i2,v2}
\fmf{dbl_wiggly, label=$h_{\mu\nu}$}{v1,v2}
\fmf{photon,label.side=left}{v1,o1}
\fmf{plain,label.side=right}{v2,o2}
\fmflabel{$A^\mu$}{i1}
\fmflabel{$A^\mu$}{o1}
\fmflabel{$\phi$}{o2}
\fmflabel{$\phi$}{i2}
\end{fmfgraph*}\nonumber
\end{equation}
\\
In this figure, the single wiggly line represents a photon corresponding to the quantum of the vector field $A^\mu$ in the action, the double wiggly line represents a graviton $h_{\mu\nu}$ propagating on a flat background spacetime, and the straight line is a propagating scalar particle associated with $\phi$. 

In order to evaluate amplitudes we need to specify some additional information, namely the momenta and helicities of the external particles. Our convention will be to have all the particles outgoing. This means that, if we label the particles in the diagram counterclockwise, say, from 1 to 4, then the momenta satisfy $p_1+p_2+p_3+p_4=0$. for concreteness, let us label the momenta of the photon legs by $p_1$ and $p_2$ and the momenta of the scalar particle legs by $p_3$ and $p_4$. The scalar has no helicity label, but the helicity of the photons $h_1$ and $h_2$ can take values $\pm1$. From the action \eqref{eq:EMSact} we can derive the Feynman rules and obtain the form of interaction vertices, propagators and polarization vectors. At this point, we recommend readers unfamiliar with the details of extracting Feynman rules from a given Lagrangian to \cite{Kumericki:2016}. Continuing, with $h_1=-1$ and $h_2=+1$, for example, we have the Feynman diagram\\
\begin{equation}
\begin{fmfgraph*}(160,60)
\fmfleft{i1,i2}
\fmfright{o1,o2}
\fmf{photon,label=$1^{-1}$,label.side=left}{i1,v1}
\fmf{plain,label=$4^0$,label.side=right}{i2,v2}
\fmf{dbl_wiggly, label=$h_{\mu\nu}$}{v1,v2}
\fmf{photon,label=$2^{+1}$,label.side=left}{v1,o1}
\fmf{plain,label=$3^0$,label.side=right}{v2,o2}
\fmflabel{$A^\mu$}{i1}
\fmflabel{$A^\mu$}{o1}
\fmflabel{$\phi$}{o2}
\fmflabel{$\phi$}{i2}
\end{fmfgraph*}
\label{fig:lb2}
\end{equation}
\\
In what follows, we will drop the explicit fields from these figures, as the information about the spin is encoded in the superscript attach to each particle number (as well as in the style of the line used in the graph). In our conventions then, this diagram is interpreted as encoding the semi-classical gravitational interaction between an incoming photon with helicity $+1$ and momentum $-p_1$ and a massive scalar particle with momentum $-p_4$, resulting in a photon with helicity $+1$ and momentum $p_2$ and a massive scalar with momentum $p_3$.

Following the procedure discussed in \cite{Kumericki:2016}, in order to write down the Feynman rules for the action (\ref{eq:EMSact}), we expand the metric $g_{\mu\nu}(x)$ in small fluctuations about flat Minkowski spacetime as 
\begin{equation}
  g_{\mu\nu} = \eta_{\mu\nu} + \kappa h_{\mu\nu},
\end{equation}
where $\kappa^2 = 32\pi G$. The perturbation $h_{\mu\nu}(x)$ now serves as the field describing the gravitational field (or gravitons, when quantized). Treating the gravitons in the same way as photons are treated in \cite{Kumericki:2016} yields the interaction vertex  
\begin{align}
  V^{\mu\nu}(1^02^0) = -\frac14 i \kappa \left[ p_1^\mu p_2^\nu + p_2^\mu p_1^\nu - \eta^{\mu\nu}  
  \left( p_1\cdot p_2 - m^2  \right) \right]\,,\label{eq:scalgravv}
\end{align}
between two scalar particles and a graviton. Similarly, the interaction vertex between two photons and a graviton is given by
\begin{align}
V^{\rho\sigma\gamma\delta}(1^{+1}2^{-1})=V^{\rho\sigma\gamma\delta}(1^{-1}2^{+1})
&= \frac14 i \kappa \left[\right. p_1\cdot p_2(\eta^{\rho\gamma}\eta^{\sigma\delta}+\eta^{\rho\delta}\eta^{\sigma\gamma}-\eta^{\rho\sigma}\eta^{\gamma\delta}) \nonumber\\
&+\eta^{\rho\sigma}p_1^{\delta}p_2^{\gamma} +\eta^{\gamma\delta}(p_1^{\rho}p_2^{\sigma}+p_1^{\sigma}p_2^{\rho})\nonumber\\
&-(\eta^{\rho\delta}p_1^{\sigma}p_2^{\gamma} +\eta^{\sigma\delta}p_1^{\rho}p_2^{\gamma}+\eta^{\sigma\gamma}p_1^{\delta}p_2^{\rho}+\eta^{\rho\gamma}p_1^{\delta}p_2^{\sigma})\left.\right],\label{eq:photgravv}
\end{align}
and vertices in which the external photons have the same helicity vanish, {\it i.e.} $V^{\rho\sigma\gamma\delta}(1^{+1}2^{+1})=V^{\rho\sigma\gamma\delta}(1^{-1}2^{-1})=0$. Since both quantities on the left hand sides of Eqs. \eqref{eq:scalgravv} and \eqref{eq:photgravv} are Lorentz tensors,  indices are raised and lowered with the Minkowski metric. For instance, $V^{\rho\sigma}_{\ \ \ \gamma\delta}(1^{+1}2^{-1})=\eta_{\gamma\mu}\eta_{\delta\nu}V^{\rho\sigma\mu\nu}(1^{+1}2^{-1})$.

When constructing the expression for the appropriate four point amplitude the vertices are stitched together by contracting their indices with those of either the graviton propagator or the appropriate polarization vectors for the external photons. To that end we need the graviton propagator. This is somewhat tediously derived from the quadratic term in the expansion of the Ricci scalar (and, for a nice treatment, we refer the reader to \cite{Zee:2003mt}). For a general momentum $q$ transferred by the graviton, the graviton propagator reads
\begin{equation}
P_{\mu\nu\alpha\beta}(q) = \frac{1}{2 q^2} \left(  \eta_{\mu\alpha}\eta_{\nu\beta}+\eta_{\mu\beta}\eta_{\nu\alpha}-\eta_{\mu\nu}\eta_{\alpha\beta} \right).\label{eq:gprop0}
\end{equation}
Putting this together then, the only two non-zero 4-point amplitudes are 
\begin{equation}
A_4(1^{+1}2^{-1}3^04^0)=
V^{\mu\nu}(3^04^0) P_{\mu\nu\rho\sigma}(q) V^{\rho\sigma}_{\ \ \ \gamma\delta}(1^{+1}2^{-1})\epsilon^{\gamma}_{+}(p_1) \epsilon^{\delta}_{-}(p_2)
\end{equation}
and
\begin{equation}
A_4(1^{-1}2^{+1}3^04^0)=
V^{\mu\nu}(3^04^0) P_{\mu\nu\rho\sigma}(q) V^{\rho\sigma}_{\ \ \ \gamma\delta}(1^{-1}2^{+1})\epsilon^{\gamma}_{-}(p_1) \epsilon^{\delta}_+(p_2).
\end{equation}
Since the polarization vectors corresponding to opposite helicities are related by complex conjugation, so too are the corresponding amplitudes, {\it i.e.}
\begin{equation}
A_4(1^{+1}2^{-1}3^04^0)^*=A_4(1^{-1}2^{+1}3^04^0).\label{eq:befcm}
\end{equation}
In order to extract a measurable quantity out of this, we will work in the center of mass frame and make the following approximations and substitutions:
\begin{itemize}
	\item First, taking the static limit for the scalar, as one would expect for a 
	massive star say, requires that we take $(p_4)_{\mu} = (p_3)_{\mu} = m \eta_{\mu 0}$.
	\item Next, we assume that the photon deflection angle, $\theta_{D}$, is small. 
	This is equivalent to the approximation of small momentum transfer, 
	or $P_{12}^2=(p_1+p_2)^2 \approx 0$.
	\item Finally, keeping the static limit and the small angle approximation in mind, we use momentum 
	conservation at the photon vertex, $P_{12}^2 = (p_1+p_2)^2 = 2 p_1 \cdot p_2$, and the fact that 
	the change in energy between the incoming and outgoing photons is small, $E_2 - E_1 \approx 0$ 
	to write $P_{12}^2 \approx -4 E_1^2 \sin^2(\theta/2)$.
\end{itemize}

With the above in mind and after some algebra taking into account the symmetries of the problem and momentum conservation, the non-zero amplitudes can be written as
\begin{equation}
A(1^{\pm1}2^{\mp1}3^04^0)=\frac{ \kappa^2 m ^2}{4 \sin^2(\theta/2)}\epsilon_{\pm}(p_1) \cdot \epsilon_{\mp}(p_2).\label{eq:aftcm}
\end{equation}
Next, we need to evaluate the differential cross-section which is a measure of the probability of interaction between the two incoming particles, and therefore proportional to the modulus squared of the scattering amplitude. This can be done for processes involving definite helicities or for unpolarized photons and it is not difficult to show that the result is the same in all these situations. In our case, where $h_1=-1$ and $h_2=+1$ as depicted in Fig. \eqref{fig:lb2}, the differential cross-section
\begin{equation}
\frac{d\sigma^{(-1,+1)}}{d\Omega}=\frac{1}{64\pi^2s_{14}}|A(1^{-1}2^{+1}3^04^0)|^2,
\end{equation}
where $s_{14}=-(p_1+p_4)^2=m^2-2p_1\cdot p_4$ is one of the Mandelstam invariants defined in Eq. \eqref{eq:Mandelstam}. In the low-energy limit, in which the photon energy is small when compared to the mass $m$ of the scalar, $s_{14}\simeq m^2$. Then, using the small angle approximation $\sin(\theta/2) \simeq \theta/2$ and the definition of the polarization vectors in Eq. \eqref{eq:polvec}, 
\begin{align}
\frac{d\sigma}{d\Omega} &=\frac{1}{128\pi^2 m^2}\bigg{|} \frac{\kappa^2 m ^2}{4 \sin^2(\theta/2)} \bigg{|}^2 \sum_{h=\pm} |\epsilon_{h}(p_1) \cdot \epsilon_{-h}(p_2)|^2           \nonumber\\
&=\frac{16 G^2 m^2}{\theta^4}.\label{eq:cross-sec}
\end{align}
To compare this with the more familiar result from GR, we need to relate the cross-section to the impact parameter $b$, the perpendicular offset of the incoming photons. Some elementary geometry shows that $\sigma = \pi b^2$, or, infinitesimally,
\begin{equation}
 b~db = -\frac{d\sigma}{d\Omega}\sin\theta d\theta.
\end{equation}
The scattering angle can be found by integrating this equation, using Eq. \eqref{eq:cross-sec} in the small angle approximation
\begin{equation}
\int b~db = \frac{b^{2}}{2} = -\int \frac{d\sigma}{d\Omega}\theta d\theta = \frac{8G^{2}m^{2}}{\theta^{2}}\,,
\end{equation}
where the integration constant can be set to zero by comparing to the flat space ($m=0$) case. Physically, we expect the maximum deflection angle $\theta_{D}$ when the photon just grazes the surface of the lens where $b=R_{0}$ and  
\begin{equation}
\theta_{\rm D}= \frac{4G m}{R_{0}}.
\end{equation}
This is nothing but the classical result for the gravitational light-bending angle that we obtained in Eq. \eqref{eq:scattangle}, if we make the natural identification between the mass of the scalar $m$ and the Schwarzschild mass $M$. \\

\newpage
Of course the reader might rightly be concerned with the fact that our computation assumes a single graviton interaction along the trajectory of the scattered photon\footnote{We would like to thank Daniel Grin for a discussion clarifying this issue and refering us to \cite{Caldwell:2006gu}.}. After all, in quantum mechanics, the philosophy that any process not expressly forbidden (by say the symmetries of the problem) must be accounted for would seem to dictate that we necessarily include so-called ladder diagrams like 
\begin{align}
&\begin{gathered}
\begin{fmfgraph*}(160,60)
\fmfleft{i1,i2}
\fmfright{o1,o2}
\fmf{photon,label.side=left}{i1,v1}
\fmf{plain,label.side=right}{i2,v2}
\fmf{dbl_wiggly}{v1,v2}
\fmf{photon}{v1,o1}
\fmf{plain}{v2,o2}
\end{fmfgraph*}\nonumber
\end{gathered}
+
\begin{gathered}
\begin{fmfgraph*}(160,60)
\fmfleft{i1,i2}
\fmfright{o1,o2}
\fmf{photon}{i1,v1,v2,o1}
\fmf{plain}{o2,v4,v3,i2}
\fmf{dbl_wiggly,tension=0.5}{v1,v3}
\fmf{dbl_wiggly,tension=0.5}{v2,v4}
\end{fmfgraph*}\nonumber
\end{gathered}
+\\\\& 
\begin{gathered}
\begin{fmfgraph*}(160,60)
\fmfleft{i1,i2}
\fmfright{o1,o2}
\fmf{photon}{i1,v1,v2,o1}
\fmf{plain}{o2,v4,v3,i2}
\fmf{dbl_wiggly,tension=0.5}{v1,v4}
\fmf{dbl_wiggly,tension=0.5}{v2,v3}
\end{fmfgraph*}\nonumber
\end{gathered}
+
\begin{gathered}
\begin{fmfgraph*}(160,60)
\fmfleft{i1,i2}
\fmfright{o1,o2}
\fmf{photon}{i1,v1,v2,v3,o1}
\fmf{plain}{o2,v4,v5,v6,i2}
\fmf{dbl_wiggly,tension=0.5}{v1,v6}
\fmf{dbl_wiggly,tension=0.5}{v2,v5}
\fmf{dbl_wiggly,tension=0.5}{v3,v4}
\end{fmfgraph*}\nonumber
\end{gathered}
+ ~\cdots
\end{align}
However, by comparing the time that a photon incident on a target with impact parameter $b$ spends near the gravitational lens, $\Delta t_{\rm near}\sim b/c$, and the gravitational interaction time $\Delta t_{\rm int}\sim b/c$ we see that the photon effectively experiences a single scattering interaction. This should be contrasted with a nonrelativistic particle moving at speed $v$ and scattering off the same massive lens. In this case while the gravitational interaction time is the same $\Delta t_{\rm int}$ as for the photon, the time spent in the effective scattering zone of the lens is much longer at $\sim b/v$. Consequently, we would expect that scattering event to be made up of a large number of single graviton exchanges with the lensing body.

As Feynman diagram computations go, this was relatively simple. Imagine instead that we want to study the scattering of a gravitational wave passing close by the sun following the same methodology as for photons. In this case we would promote the two photons in our Feynman diagram to gravitons. Since we already have the two-scalar graviton vertex and the graviton propagator we only need the three graviton vertex. This derives simply from the cubic term in the $h_{\mu\nu}$-expansion of the Einstein-Hilbert action as
\begin{equation}\label{gvertex}{
	\begin{aligned}
	{V}_{\alpha  \beta \gamma \delta }^{\mu
		\nu}(1^{\pm2}2^{\pm2})&=-\frac{i\kappa}2 \bigg({\cal P}_{\alpha \beta
		\gamma \delta }\bigg[{p_1}^\mu {p_1}^\nu+ ({p_1}-{p_2})^\mu ({p_1}-{p_2})^\nu +{p_2}^\mu
	{p_2}^\nu-
	\frac32\eta^{\mu \nu}(p_2)^2\bigg]\\[0.00cm]&
	+2{p_2}_\lambda {p_2}_\sigma\bigg[ I_{\alpha \beta }^{\ \ \
		\sigma\lambda}I_{\gamma \delta }^{\ \ \ \mu \nu} + I_{\gamma
		\delta }^{\ \ \ \sigma\lambda}I_{\alpha \beta }^{\ \ \ \mu \nu}
	-I_{\alpha \beta }^{\ \ \ \mu  \sigma} I_{\gamma \delta }^{\ \ \
		\nu \lambda} - I_{\gamma \delta }^{\ \ \ \mu \sigma} I_{\alpha
		\beta }^{\ \ \ \nu \lambda}
	\bigg]\\[0cm]&
	+\bigg[{p_2}_\lambda {p_2}^\mu \bigg(\eta_{\alpha \beta }I_{\gamma \delta
	}^{\ \ \ \nu \lambda}+\eta_{\gamma \delta }I_{\alpha \beta }^{\ \
	\ \nu \lambda}\bigg) +{p_2}_\lambda {p_2}^\nu \left(\eta_{\alpha \beta
}I_{\gamma \delta }^{\ \ \ \mu \lambda}+\eta_{\gamma \delta
}I_{\alpha \beta }^{\ \ \ \mu  \lambda}\right)\\&
-(p_2)^2\left(\eta_{\alpha \beta }I_{\gamma \delta }^{\ \ \ \mu
	\nu}-\eta_{\gamma \delta }I_{\alpha \beta }^{\ \ \ \mu \nu}\right)
-\eta^{\mu \nu}{p_2}_\sigma {p_2}_\lambda\left(\eta_{\alpha \beta
}I_{\gamma \delta }^{\ \ \ \sigma\lambda} +\eta_{\gamma \delta
}I_{\alpha \beta }^{\ \ \
\sigma\lambda}\right)\bigg]\\[0cm]&
+\bigg[2{p_2}_\lambda\big(I_{\alpha \beta }^{\ \ \
	\lambda\sigma}I_{\gamma \delta \sigma}^{\ \ \ \ \nu}({p_1}-{p_2})^\mu
+I_{\alpha \beta }^{\ \ \ \lambda\sigma}I_{\gamma \delta
	\sigma}^{\ \ \ \ \mu }({p_1}-{p_2})^\nu -I_{\gamma \delta }^{\ \ \
	\lambda\sigma}I_{\alpha \beta \sigma}^{\ \ \ \ \nu}{p_1}^\mu
-I_{\gamma \delta }^{\ \ \ \lambda\sigma}I_{\alpha \beta
	\sigma}^{\ \ \ \ \mu }{p_1}^\nu \big)\\& +(p_2)^2\left(I_{\alpha \beta
	\sigma}^{\ \ \ \ \mu }I_{\gamma \delta }^{\ \ \ \nu \sigma} +
I_{\alpha \beta }^{\ \ \ \nu \sigma}I_{\gamma \delta \sigma}^{\ \
	\ \ \mu }\right) +\eta^{\mu \nu}{p_2}_\sigma {p_2}_\lambda\left(I_{\alpha
	\beta }^{\ \ \ \lambda\rho}I_{\gamma \delta  \rho}^{\ \ \ \
	\sigma} +I_{\gamma \delta }^{\ \ \ \lambda\rho}I_{\alpha \beta
	\rho}^{\ \ \ \
	\sigma}\right)\bigg]\\[0cm]&
+\bigg\{[(p_1)^2+({p_1}-{p_2})^2]\big[I_{\alpha \beta }^{\ \ \ \mu
	\sigma}I_{\gamma \delta \sigma}^{\ \ \ \ \nu} +I_{\gamma \delta
}^{\ \ \ \mu  \sigma}I_{\alpha \beta \sigma}^{\ \ \ \ \nu}
-\frac12\eta^{\mu \nu}{\cal P}_{\alpha \beta \gamma \delta
}\big]\\&-\left(I_{\gamma \delta }^{\ \ \ \mu \nu}\eta_{\alpha
\beta }(p_1)^2+I_{\alpha \beta }^{\ \ \ \mu \nu}\eta_{\gamma \delta
}({p_1}-{p_2})^2\right)\bigg\}\bigg),
\end{aligned}}
\end{equation}
with ${\cal P}_{\mu\nu\alpha\beta}=\left(  \eta_{\mu\alpha}\eta_{\nu\beta}+\eta_{\mu\beta}\eta_{\nu\beta}-\eta_{\mu\nu}\eta_{\alpha\beta} \right)/2$  and $I_{\mu\nu\alpha\beta} =\left(  \eta_{\mu\alpha}\eta_{\nu\beta}+\eta_{\mu\beta}\eta_{\nu\beta}\right)/2$.
The partial amplitude is then written as
\begin{equation}
A(1^{\pm2}2^{\pm2}3^04^0)
=
V^{\mu\nu}(3^{0}4^{0}) P_{\mu\nu\rho\sigma}(q) V^{\rho\sigma\gamma\delta\alpha\beta}(1^{\pm2}2^{\pm2}) \epsilon_{\gamma}^{\pm}(p_1) \epsilon_{\delta}^{\pm}(p_1)  \epsilon_{\gamma}^{\pm}(p_2) \epsilon_{\delta}^{\pm}(p_2).
\end{equation}
In principle this looks straightforward enough. However, it should be clear by the proliferation of terms that now need to be evaluated in order to compute the scattering amplitude that there is a pressing need for a more efficient formalism to deal with problems like this. 

\section{Modern amplitude methods: BCFW relations}
\subsection{Spinor-helicity formalism}
 
Like any new set of tools, the techniques we are about to introduce will take some getting used to. We will keep the presentation deliberately pragmatic and focused toward solving a particular problem in order that the reader can look back and compare with the foregoing sections. And while we will assume some level of familiarity with group theory and the basics of quantum field theory on the part of the reader, we would advise a digression to the appendices at this point, if only to get acquainted with our notation. This section closely follows the treatment found in \cite{Elvang:2013cua}, \cite{Carrasco:2015iwa} and \cite{Dixon:2013uaa}, and which we refer the interested reader for further details. Even though many of these techniques were borne of superstring theory, their utility is primarily in the ``real world" of 4-dimensional spacetime which is, unless explicitly stated otherwise, where we will remain. 

The key point that we start from is that in four spacetime dimensions, the familiar Lorentz group of  rotations and boosts can be mapped to the group of $2\times 2$ matrices with complex entries and unit determinant,
\begin{equation}\label{key}
SO(1,3)\simeq SL(2,\C).
\end{equation}
This innocuous observation allows us to decompose any Lorentz four vector\footnote{Our notation is looking forward to the role that $P^{\mu}$ will play as a 4-momentum but, for now, it really can be any old Lorentz vector.}, $P^{\mu}$, into a so-called {\bf bi-spinor}, a $2\times 2$ matrix with two indices from $SL(2,\C)$, {\it i.e.}
\begin{equation}
P^\mu \in SO(1,3)\ \longrightarrow\ P_{a\dot{b}} = P_\mu(\sigma^\mu)_{a\dot{b}} \in SL(2,\C).\label{eq:isodef}
\end{equation}
This mapping is defined through the set of matrices $(\sigma^\mu)_{a\dot{b}}$, which consists of the identity $\mathbf{1}_{a\dot{b}}$ and the usual Pauli matrices of quantum mechanics (see, for example. eq. \eqref{paulim}),
\begin{equation}
(\sigma^\mu)_{a\dot{b}}=(\mathbf{1},\sigma^i)_{a\dot{b}}. 
\end{equation}
Notice that since the Lorentz indices in \eqref{eq:isodef} are summed over, we have effectively traded {\it one} spacetime index for {\it two} matrix indices with the undotted index corresponds to the row label and the dotted one to the column label. In this language, the invariant square $P_\mu P^\mu$ of the vector $P^\mu$ is given by the determinant of the bi-spinor $P_{a\dot{b}}$. For null vectors, this means that the $2\times 2$ matrix $P_{a\dot{b}}$ can be written as the product of two-component spinors since,
\begin{equation}
\det(P_{a\dot{b}}) = 0\ \Longleftrightarrow\ P_{a\dot{b}} = -\lambda_a\tilde{\lambda}_{\dot{b}}.
\end{equation}
These objects are complex columns known as {\bf Weyl spinors}, and they will provide the basic building blocks for what follows. They will obviously depend on the real vector $P^\mu$, and have components:
\begin{equation}
\lambda_a = \begin{pmatrix}
\sqrt{P^0 + P^3} \\ \frac{\displaystyle P^1 + iP^2}{\sqrt{\displaystyle P^0 + P^3}}
\end{pmatrix},\qquad\qquad\tilde{\lambda}_{\dot{a}} = \begin{pmatrix}
\sqrt{P^0 + P^3} \\ \frac{\displaystyle P^1 - iP^2}{\sqrt{\displaystyle P^0 + P^3}}
\end{pmatrix}.
\end{equation}
Notice that for a real vector, these associated spinors are complex conjugates of one another. This is no longer the case if the components of $P^{\mu}$ are complex. Instead of using Eq. \eqref{eq:isodef}, the mapping between Lorentz vectors and bi-spinors may alternatively be defined 
through
\begin{equation}
P^{\dot{a}b}\equiv P_\mu({\bar{\sigma}}^{\mu})^{\dot{a}b},
\end{equation}
where now
\begin{equation}
(\bar{\sigma}^\mu)^{\dot{a}b}=(\mathbf{1},-\sigma^i)^{\dot{a}b}. 
\end{equation}
Technically speaking, this furnishes another representation of the bi-spinors. However, these two representations can be linked with the completely antisymmetric Levi-Civita symbols $\epsilon^{ab}$ and $\epsilon_{ab}$ defined as $\epsilon^{12} = -\epsilon^{21} = \epsilon_{21} = -\epsilon_{12} = 1$, with all the other components zero. Specifically since,
\begin{equation}
(\bar{\sigma}^\mu)^{\dot{a}b}=\epsilon^{bc}\epsilon^{\dot{a}\dot{d}}(\sigma^\mu)_{c\dot{d}}\,.
\end{equation}
it follows that the bi-spinors $P_{a\dot{b}}$ and $P^{\dot{a}b}$ are not independent quantities. Raising and lowering spinor indices is also naturally defined for Weyl spinors so that, if $P^{\dot{a}b}=-\tilde{\lambda}^{\dot{a}}\lambda^b$,
\begin{equation}
\lambda_a = \epsilon_{ab}\lambda^b,\qquad\tilde{\lambda}_{\dot{a}} = \epsilon_{\dot{a}\dot{b}}\tilde{\lambda}^{\dot{b}}.\label{eq:updown}
\end{equation}
%

\subsubsection{Square and angle bra-ket notation \label{sec:braket}}

To circumvent a proliferation of unwieldy dotted and undotted indices, we will now introduce a simple and intuitive modern notation for our two-component spinors. Let's consider the Lorentz 4-momentum $p^{\mu}$ for definiteness sake and carry out the split into Weyl spinors above. then associated to this vector, we will have the spinors\footnote{We find the following mnemonics helpful: (bra)ket $\leftrightarrow$ (anti-) spinor and (square) angle $\leftrightarrow$ (positive) negative helicity.}
\begin{itemize}
	\item $\lambda_a = |p]_a \equiv |p]~~~~~$ A positive helicity spinor\\
	\item $\tilde{\lambda}^{\dot{a}} = \ket{p}^{\dot{a}} \equiv \ket{p}~~~~$ A negative helicity spinor\\
	\item $\lambda^a = [p|^a\equiv [p|~~~~~$ A positive helicity anti-spinor\\
	\item $\tilde{\lambda}_{\dot{a}} = \bra{p}_{\dot{a}}\equiv \bra{p}~~~~$ A negative helicity anti-spinor
\end{itemize}
All of the angle and square spinors commute with one another, and the operations of raising and lowering indices defined in Eq. \eqref{eq:updown} now take the form,
\begin{equation}
[p|^a\,=\, \epsilon^{ab} |p]_b\qquad\ket{p}^{\dot{a}} \,=\, \epsilon^{\dot{a}\dot{b}}  \bra{p}_{\dot{b}}\, .
\end{equation}
In words, the Levi-Civita symbol converts a spinor into an anti-spinor while preserving its helicity. Moreover, for real momenta $p^\mu$, complex conjugation changes spinors into anti-spinors and flips their helicities since, as can be checked explicitly,
\begin{equation}
[p|^a = (\ket{p}^{\dot{a}})^*,\qquad \bra{p}_{\dot{a}} = (|p]_a)^*.
\end{equation}
The four-momentum of some particle in our new notation is again the product of two spinors of opposite helicity,
\begin{equation}
p_{a\dot{b}}~=~ -|p]_a\, \bra{p}_{\dot{b}} \, ,~~~~~~~\label{eq:1decomp}
p^{\dot{a}b}~=~- \ket{p}^{\dot{a}}\, [p|^{b}  \, .~~~~~~~
\end{equation}
It will help to illustrate these concepts with a concrete example.
\begin{eBox}
Let's pick a specific null four-vector (which may or may not correspond to a momentum),
\begin{equation}
p^\mu = (E,E\sin\theta\cos\varphi, E\sin\theta\sin\varphi,E\cos\theta).
\end{equation}
The mapping in eq. \eqref{eq:isodef} maps this vector to a bi-spinor:
\begin{align}
\left(p_\mu\sigma^\mu\right)_{a\dot{b}} &= E\begin{pmatrix}\cos\theta - 1  & \sin\theta\cos\varphi -i\sin\theta\sin\varphi \\ \sin\theta\cos\varphi +i\sin\theta\sin\varphi & -(\cos\theta + 1)\end{pmatrix}\,.
\end{align}
This in turn can be written, with a little trigonometric gymnastics, in the more convenient form
\begin{align}
\left(p_\mu\sigma^\mu\right)_{a\dot{b}} &= E\begin{pmatrix}\cos\theta - 1  & \sin\theta\,e^{-i\varphi} \\ \sin\theta\,e^{i\varphi} & -(\cos\theta + 1)\end{pmatrix}\\
&= 2E\begin{pmatrix}-\sin^2(\theta/2)  & \sin(\theta/2)\cos(\theta/2)\,e^{-i\varphi} \\  \sin(\theta/2)\cos(\theta/2)\,e^{i\varphi} & -\cos^2(\theta/2)\end{pmatrix}.
\end{align}
The determinant of this matrix is clearly identically zero, so we can now happily write it as the (outer) product of two vectors:
\begin{equation}
-\spvec{a~b}\otimes\begin{pmatrix}c \\ d\end{pmatrix} = -\begin{pmatrix}ac  & ad \\ bc & bd\end{pmatrix} = \begin{pmatrix}-\sin^2(\theta/2)  & \sin(\theta/2)\cos(\theta/2)\,e^{-i\varphi} \\  \sin(\theta/2)\cos(\theta/2)\,e^{i\varphi} & -\cos^2(\theta/2)\end{pmatrix}.
\end{equation}
With four equations in four unknowns, this equation is solved by choosing
\begin{equation}
a = c = \sin(\theta/2),\qquad b = d^* = -\cos(\theta/2)\,e^{i\varphi}.
\end{equation}
Finally then, we identify the spinors
\begin{equation}
|p]_a = \sqrt{2E}\spvec{\sin(\theta/2),-\cos(\theta/2)\,e^{i\varphi}},\qquad \bra{p}_{\dot{b}} = \sqrt{2E}\begin{pmatrix}\sin(\theta/2) \\ -\cos(\theta/2)\,e^{-i\varphi}\end{pmatrix}\,. 
\end{equation}
The pre-factors in each of these expressions have been inserted to ensure that the spinors are dimensionally correct. A similar procedure can be followed to show the equivalent result for $p^{\dot{a}b}$.
\end{eBox}
Using these explicit expressions for the momentum spinors, it is interesting to note that acting on either of them with the original bispinor $p_{a\dot{b}}$ gives us zero; for instance,
\begin{align}
p_{a\dot{b}}\ket{p}^{\dot{b}} &= p_{a\dot{b}}\epsilon^{\dot{b}\dot{c}}\bra{p}_{\dot{c}}\\
&= \sqrt{2E}\begin{pmatrix}-\sin^2(\theta/2)  & \sin(\theta/2)\cos(\theta/2)\, e^{-i\varphi} \\  \sin(\theta/2)\cos(\theta/2) e^{i\varphi} & -\cos^2(\theta/2)\end{pmatrix}\begin{pmatrix}0  & 1 \\ -1 & 0\end{pmatrix}\begin{pmatrix}\sin(\theta/2) \\ -\cos(\theta/2)\,e^{-i\varphi}\end{pmatrix}\\
&= \begin{pmatrix}-\sin^2(\theta/2)\cos(\theta/2)\,e^{-i\varphi} + \sin^2(\theta/2)\cos(\theta/2)\,e^{-i\varphi} \\ \cos^2(\theta/2)\sin(\theta/2) - \cos^2(\theta/2)\sin(\theta/2) \end{pmatrix}\nonumber\\
&=0.
\end{align}
This is one form of the massless {\bf Weyl equation}, which is the two-component spinor equivalent of the massless Dirac equation. This equation is satisfied by all the momentum spinors introduced above, and will be extremely useful in order to deal with scattering amplitudes. To conclude this example, we will summarise the form of the Weyl equation for the different momentum spinors,
\begin{equation}
p_{a\dot{b}}\ket{p}^{\dot{b}} = 0,\qquad\bra{p}_{\dot{a}}p_{\dot{a}b} = 0,\qquad[p|^a p_{a\dot{b}} = 0,\qquad p^{\dot{a}b}[p|_b = 0.\label{eq:weyleq}
\end{equation}

Continuing on with our discussion of the bra-ket notation, it is worth highlighting several other relevant properties of spinors that are captured by this notation. The first of these is the angle spinor bra-ket and square spinor bra-ket which, for two lightlike vectors $p^\mu$ and $q^\mu$, are defined as
\begin{equation}
\braket{pq} \,=\,{\bra{p}}_{\dot{a}}\, \ket{q}^{\dot{a}} = -\braket{qp\,},~~~~~~~~
[ p \, q ]  \,=\, [p|^{a} \,|q]_a = -[ q \, p ]\,.
\end{equation}
The antisymmetry of the bra-kets follows from the Levi-Civita symbols that are used to raise and lower spinor indices. For the same reason, all other combinations of bra-kets {\it e.g.} $\langle pq]$ vanish. These new {\bf spinor-helicity variables} satisfy a number of remarkable identities, many of which have been enumerated in detail in, for example, appendix A of \cite{Elvang:2013cua}. While we will not recount all of them here, it will be useful for our purposes to elaborate on one or two. The first of these is that  two null vectors $p^{\mu}$ and $q^{\mu}$ satisfy
\begin{equation}
  (p+q)^2 = 2p\cdot q = \braket{pq}[pq]\,.\label{eq:viprel}
\end{equation}
The second is the re-formulation of momentum conservation in spinor-helicity variables. If all the external particles (corresponding to external lines in the Feynman diagram) are massless then, starting from the momentum conservation equation \eqref{eq:momcons} and multiplying both sides from the left and right with $\bra{q}$ and $|k]$ respectively gives
\begin{align}
  0 &= \bra{q}\sum_{i=1}^n p_i |k]\nonumber\\
  &= \sum_{i=1}^n \langle q|p_i|k]\nonumber\\
  &= \sum_{i=1}^n \langle q|_{\dot{a}}\, p_{i}^{\dot{a}b} \,|k]_b\nonumber\\
  &= \sum_{i=1}^n \langle q|_{\dot{a}}\, ( -\ket{p_{i}}^{\dot{a}} [p_{i}|^b )  \,|k]_b\nonumber\\
  &=-\sum_{i=1}^n \braket{qi}[ik].\label{eq:mc0}
\end{align}
In the last equality, we have introduced the shorthand notation which replaces the internal momenta 
$p_i$ with $i$, so that $p_i =-\ket{i}[i|$. This notation will be used frequently in what follows.
Eq. \eqref{eq:mc0} leads to a new definition of momentum conservation, valid for \textit{any} $q$ and $k$,
\begin{equation}
  \sum_{i=1}^n \braket{qi}[ik] = 0.
\end{equation}
The final quantity that requires introduction is the {\bf polarization vector}, since we are going to be dealing with massless particles that have spin. Following the conventions laid out in \cite{Elvang:2013cua}, the positive and negative helicity polarization vectors are written as 
\begin{equation}
  \epsilon^\mu_+(p;q) = -\frac{\bra{q}\gamma^\mu|p]}{\sqrt{2}\braket{qp}},
  \qquad\qquad\epsilon^\mu_-(p;q) = -\frac{\bra{p}\gamma^\mu|q]}{\sqrt{2}[qp]}\,,\label{eq:polvec}
\end{equation}
respectively. Note that these two vectors are related by complex conjugation:
\begin{equation}
[\epsilon^\mu_-(p;q)]^*=-\epsilon^\mu_+(p;q).\label{eq:polvecqq}
\end{equation} 

Let's unpack these objects. The $\gamma$-matrices satisfy the anticommutation relations $\left\{\gamma^{\mu},\gamma^{\nu}\right\} 
= -2\eta^{\mu\nu}$ and can be realized in terms of the Pauli matrices as
\begin{eqnarray}
  \gamma^{\mu} = \left(
                        \begin{array}{cc}
                           0 & \left(\sigma^{\mu}\right)_{a\dot{b}}\\
                          \left({\bar\sigma}^{\mu}\right)^{\dot{a}b} & 0 
                        \end{array}
                       \right)\,.
\end{eqnarray} 
The polarization vector is a function of the momentum $p$ and an arbitrary reference momentum $q$, which is an \textit{auxiliary} variable in that it does not correspond to any physical quantity. Its presence reflects a gauge freedom in the formalism; we are free to shift the polarization vectors by an arbitrary constant multiple of the momentum $p$ without changing the on-shell amplitude\footnote{Technically, this is encoded in the {\bf Ward identity}, $p_{\mu}A_{n}^{\mu}=0$.}, $A_{n}$. We can freely choose $q$ to be whatever we like, however it is often useful to make $q$ equal to one of the external momenta $p_i$ when actually trying to calculate amplitudes. It is important that the final form of the amplitude \textit{does not} depend on this choice of momentum however. In practice, the choice does tend to matter and making clever choices can often greatly simplify calculations.

Our focus in this article is to use recursion relations to build up amplitudes relevant to astrophysics. At tree level ({\it i.e.} without quantum corrections), there are only two ingredients that are needed in order to calculate anything we might want: 3-point amplitudes, which represent interactions of three particles, and 2-point functions (Green's functions or propagators), which represent a single particle propagating between two spacetime points.

To derive the $n$-point amplitudes that encode the scattering of $n$ particles we would need to start from the action of the theory and derive the Feynman rules. For many theories like Einstein gravity, these often result in a hideous mess of many terms and a plethora of indices, even though the on-shell final expression is often ludicrously simple by comparison (just look at the vertex in Eq. \eqref{gvertex} above for a reminder). The source of this complication is more often than not, the aforementioned virtual particles which are nothing more than a convenient fiction that we first introduce and then remove, purely to make locality manifest. An alternative and much simpler approach is to use the symmetries of the problem to constrain the possible final answers, never introducing the notion of virtual particles at all. This will be the subject of the next section.

\subsubsection{Little group scaling \label{sec:lgs}}
Pound for pound of effort, dimensional analysis is one of the most powerful qualitative tools in physics \cite{Migdal:1977bq}. It turns out that it is no less useful here, where it goes by the name of little group scaling. The {\bf little group} is the subgroup of some group that leaves a particular state invariant. Specifically, if the group $G$ acts on a space $M$ in which $m\in M$ is some fixed element   and $H \subset G$ is a subgroup that acts on $m$ leaving it invariant then $H$ is called a little group of $G$.

For our purposes, the group of interest will be the Poincar\'e group of spatial rotations, boosts and spacetime translations in four dimensions. This group acts on the space of 4-vectors $x^{\mu}$. 
If $x^{\mu}$ is a timelike vector then the little group is the $SO(3)\simeq SU(2)$ subgroup of the Poincar\'e group in the 3-space orthogonal to $x^{\mu}$. On the other hand, if $x^{\mu}$ is 
spacelike or null, the little group will be $SO(2,1)$ or $SO(2)\simeq U(1)$ respectively.

In introducing the spinor helicity formalism, we used the central notion of a null vector being represented as a momentum bi-spinor, $p_{a\dot{b}}=  -|p]_a \bra{p}_{\db}$. For real momenta, if we take $t$ to be a complex phase then it is clear that scaling the individual spinors 
\begin{equation}
  \ket{p}\rightarrow t\ket{p},\qquad\qquad |p]\rightarrow t^{-1}|p],
\end{equation}
keeps the momentum bi-spinor invariant. For complex momenta, any complex number $t$ will do as the little group scale factor. Physically, only external momenta scale under the little group; vertices and internal lines do not.

Suppose we have two incoming massless particles with momenta $p_1$ and $p_2$ that interact to produce a single massless particle with momentum $p_3$. To state clearly the following argument we will briefly permit these momenta to be complex (more details about complex extensions will be given in section \ref{sec:cshifts}). Energy-momentum conservation demands that $p_1 + p_2 + p_3 = 0$, and that $(p_1 + p_2)^2 = p_3^2 = 0$. In spinor helicity notation, this means that $\braket{12}[12]$ = 0. Since the momenta $p_1$ and $p_2$ are complex, the quantities $\langle12\rangle$ and $[12]$ are independent. Let's suppose further that $[12]$ is zero in order to make the product zero. In that case, we must also have $\braket{12}[23] = \bra{1}(p_1 + p_3)\ket{3} = 0$, so $[23] = 0$ and similarly for $[13]$. We see then that all square brackets are vanish if even one of them is zero. We could well have taken $\braket{12} = 0$ instead, but we would have simply found the same thing; all angle brackets vanish and our amplitude only depends on square brackets.

In short, this means that 3-point functions of massless particles with \emph{complex} momenta can \textit{only} depend on either angle brackets or square brackets, but not both.

A corollary of this result is that 3-point amplitudes for massless particles with \emph{real} momenta must actually vanish on-shell in 4 dimensions. Indeed, Eq. $\braket{12}[12] = 0$ implies that both $\langle12\rangle$ and $[12]$ have to be identically zero, and something similar can be shown to occur with the remaining brackets. That we have chosen complex momenta to construct non-vanishing 3-point functions may seem daft at the moment, but it will become clear in the next section. Let us also mention that there are, in principle, many ways to construct non-vanishing 3-point functions by relaxing the on-shell constraints, {\it i.e.} instead of continuing the external momenta from real to complex values, we could well consider relaxing momentum conservation instead.

Now, here's the kicker: all massless 3-particle amplitudes are completely fixed by little group scaling. The reasoning is as simple as solving three algebraic equations which can be found in \cite{Elvang:2013cua}. We will content ourselves to quote the result. The amplitudes for three massless particles with momenta and helicity $(p_{i},h_{i})$ (for $i=1,2,3$) are given by
\begin{equation}\label{3pt1}
A_3(1^{h_1}2^{h_2}3^{h_3}) = C\braket{12}^{h_3-h_1-h_2}\braket{13}^{h_2-h_1-h_3}\braket{23}^{h_1-h_2-h_3},
\end{equation}
which is valid for $ h_1 + h_2 + h_3 < 0$, or
\begin{equation}\label{3pt2}
A_3(1^{h_1}2^{h_2}3^{h_3}) = \frac{C}{[12]^{h_3-h_1-h_2}[13]^{h_2-h_1-h_3}[23]^{h_1-h_2-h_3}},
\end{equation}
valid for $h_1 + h_2 + h_3 > 0$.

In order to distinguish between an expression depending just on angle brackets or just on square brackets is a matter of locality and dimensional analysis:
\begin{quote}
	\textbf{An $n$-particle amplitude in $d=4$ must have mass-dimension $4-n$.}
\end{quote}
The constants in the above expressions are usually fixed by expanding the Lagrangian for the theory and identifying the coupling constants in the appropriate interaction term. So far our discussion has been limited to massless particle scattering. For massive particles, it is also possible to use the little group, $SU(2)$, to constrain the form of the function, however using current methods, this is accomplished at the cost of introducing additional degrees of freedom that must later be removed. 

\begin{eBox}
Let's look at an example 3-point calculation, where we will consider 3 gravitons interacting with one another, each having \textit{complex} momentum. Gravitons $h_{\mu\nu}$ are particles with helicity $h = \pm 2$, so lets first consider the simplest case where each graviton has the same helicity, say $h_1 = h_2 = h_3 = -2$. According to the formulas above, this means our amplitude must look like:
\begin{equation}
  A(1^{-2}2^{-2}3^{-2}) = C\braket{12}^{2}\braket{13}^{2}\braket{23}^{2}
\end{equation}
We now need to determine the value of $c$ from the Lagrangian. In our case, the relevant term comes from perturbatively expanding the Einstein-Hilbert action around flat space $g_{\mu\nu} = \eta_{\mu\nu} + \kappa h_{\mu\nu}$ and looking for terms that involve 3 gravitons interacting. 

Schematically, in 4-dimensions the Einstein-Hilbert action expands out as
\begin{eqnarray}
  S_{\mathrm{EH}} = \frac{1}{2\kappa^{2}}\int\,d^{4}x\,\sqrt{-g}R = \int\,d^{4}x\,
  \left(h\partial^{2}h + \kappa h^{2}\partial^{2}h + \kappa^{2}h^{3}\partial^{2}h + \ldots\right) \,,
\end{eqnarray}
where $\kappa^{2} = 8\pi G_{N}$ and the infinite series arises from inverting the metric in computing the curvature $R$ as well as an expansion of the square root. Of all these, the only term that interests us for 3-graviton scattering is the one with three factors of the graviton, $\kappa\,h\partial^{2}h$. Of course, this is a schematic representation of a whole lot more terms (see eq. \eqref{gvertex} for a gruesome reminder). For this computation though, we only really care about dimensions and the coupling constant. In natural units ($\hbar = c=1$) the action is dimensionless, so that the terms in the Lagrangian  all have mass-dimension 4. Each derivative contributes mass dimension $[\partial] = +1$, whereas the effective gravitional coupling has dimension $[\kappa] = -1$ which leaves $[h_{\mu\nu}] = +1$.
	
Looking at the form of the amplitude derived from the little group above, we realise that it has mass dimension [C] + 6, noting from eq. \ref{eq:viprel} that both angle and square brackets have mass dimension 1. We require that the entire amplitude has mass dimension 1 and therefore that [C] = -5. Clearly there is no such term in the action that corresponds to such a coupling, therefore no such amplitude can exist in this theory: same helicity gravitons don't scatter in Einstein gravity. We could have instead looked at all positive helicities, in which case we would have found that we needed a coupling with dimension [C] = +7, which also doesn't exist in the theory.
	
Now let's look at the next simplest case, where one of the particles has the opposite helicity, $h_3 = +2, h_1 = h_2 = -2$. Our amplitude now looks like:
	\begin{equation}
	A(1^{-2}2^{-2}3^{+2}) = \frac{C\braket{12}^6}{\braket{13}^2\braket{23}^2}
	\end{equation}
	This has mass dimension $[C] + 2$, meaning we need $[C] = -1$. This exactly corresponds to the form of the terms in our perturbative expansion and thus we have found a valid amplitude:
	\begin{equation}\label{key}
		A(1^{-2}2^{-2}3^{+2}) = \kappa\left(\frac{\braket{12}^3}{\braket{13}\braket{23}}\right)^2
	\end{equation}
Not only is this amplitude startlingly simple, the fact that it can be written as a perfect square is remarkable. It is the first hint of the \emph{factorization} of gravity amplitudes into a product (in this case, of gluon amplitudes). For some perspective, we invite the reader to check the computation the old fashioned way, using the Feynman rules described in \cite{Kumericki:2016}, or your favourite QFT textbook. 
\end{eBox}
\subsection{BCFW Recursion relations}

We have discussed just above the simplest scattering processes, involving only three particles. Mathematically, these are captured by the 3-point scattering amplitudes. However, most interesting physical processes involve a larger number of interacting particles, represented by $n$-point scattering amplitudes (for $n$ particles). In this section we show how to deal with these higher-point amplitudes using modern methods and, in particular, how to construct them recursively from lower-point amplitudes. We focus our discussion on the so-called Britto-Cachazo-Feng-Witten (BCFW in the following) recursion relations \cite{Britto:2004ap,Britto:2005fq,Feng:2011np}, though other constructions exist that are useful for specific situations \cite{Bedford:2005yy,Cachazo:2005ca,Cachazo:2004kj}.

In the traditional approach to perturbative quantum field theory, $n$-point scattering processes generally include the exchange of virtual particles, where the adjective virtual means that these intermediate particles cannot be detected in experiments and hence lack a physical interpretation by themselves. Virtual particles are internal, in the sense that they do not form part of the external data on which scattering amplitudes are defined, and they are off-shell, i.e. do not satisfy $p^2 = -m^2$. For instance, we discussed a particular case in section \ref{sec:gravscal} describing the exchange of a virtual graviton between a photon and a massive scalar, using the corresponding propagator. In this approach, scattering amplitudes have to be evaluated independently for every value of $n$, with increasing difficulty once we start to consider amplitudes involving more particles. The tools needed are usually condensed in the form of Feynman rules, which can be derived from the perturbative expansion of the Lagrangian of a given theory.

Modern methods provide an alternative representation of these amplitudes, in which even the internal particles being exchanged by interacting particles are on-shell. This is only possible by relaxing some of the other mathematical constraints imposed in the kinematic structure of the particles and their momenta; in particular, we introduced the notion of complex-valued momenta during the discussion of 3-point amplitudes. Using complex momenta may seem counter-intuitive at first, but this is just a convenient tool which permits us to enforce the on-shell constraint on all the particles (both external and internal) in intermediate calculations; of course, at the end of the day the physical amplitudes obtained will display dependence only on real momenta. As we discuss in the next sub-section, working on-shell has some distinct advantages. In particular, $n$-point amplitudes in a large class of theories factorize into products of lower-point on-shell amplitudes. This in turn permits us to address the evaluation of scattering amplitudes in a recursive way.

These modern methods hence exploit the power of complex analysis in order to elucidate the structure of $n$-point amplitudes. It will therefore make sense to start by developing some of the basics of complex analysis that will be needed. This section is the most technical of the paper and some readers may want to skip ahead to the punchline in section 4 on a first pass, and return with some coffee.

\subsubsection{Complex shifts and poles \label{sec:cshifts}}

An arbitrary $n$-point scattering amplitude is a function of the (real) external momenta of the interacting particles. Typically, amplitudes contain the following momentum dependence in the denominator, coming from the propagators representing virtual particles:
\begin{equation}
\frac{1}{P_{abc\cdots}^2}= \frac{1}{(p_a + p_b + p_c+\cdots)^2}.\label{eq:mnum}
\end{equation}
Here $a,b,c,...$ are indices labelling different external momenta. In the cases we will consider explicitly, only two momenta will enter in this kind of expressions, so for simplicity we will just consider internal momenta of the form $P_{ab}$ (in any case there is no difference in the following treatment, so this is just a notational issue). To be more specific, let us consider the amplitude evaluated in section \ref{sec:lbqft}. This amplitude shows this kind of dependence where $q=p_1+p_2=P_{12}$ is the sum of the momenta of the two external photons, which enters through the graviton propagator in Eq. \eqref{eq:gprop0}.

The reason we focus our attention on this feature is that an amplitude containing a dependence like Eq. \eqref{eq:mnum} would become singular if we enforce the internal momentum $P_{ab}$ of the corresponding virtual particle to be on-shell, namely $P_{ab}^2=0$. In the standard approach to quantum field theory, virtual particles are off-shell; putting the virtual particles on-shell would imply constraints on the external particles. Let us consider again the amplitude evaluated in section \ref{sec:lbqft} to illustrate this point. In this case, the momentum $q=P_{12}$ becomes on-shell only if $p_1\cdot p_2=0$, namely if the incoming and outgoing photons are orthogonal. For other values of the angle between the two photons, the momentum of the virtual particle cannot be on-shell.

However, there is a way to enforce the on-shell nature of the internal momentum $P_{ab}$, at the price of considering complex momenta. The complex extension of the momenta of the virtual particle is denoted by $\hat{P}_{ab}$ in order to stress the difference with the real momenta $P_{ab}$; let us also write $\hat{P}_{ab}=P_{ab}+\Delta P_{ab}$. More accurately, the quantity that will go on-shell is this complex extension. Due to the additional freedom associated with the introduction of the complex piece $\Delta P_{ab}$, it is possible to do so without imposing constraints on the real momenta of external particles, but rather by fixing the value of the complex part of the internal momenta. In other words, $\Delta P_{ab}$ is fixed in order to guarantee that $\hat{P}_{ab}$ is on-shell. Provided the scattering amplitudes are analytic functions, this procedure turns the scattering amplitude into a meromorphic function - analytic everywhere except at some isolated poles, namely the places in which the complex momentum $\hat{P}_{ab}$ goes on-shell. Even if this is hardly apparent at this point of the discussion, as discussed below this permits us to exploit the powerful methods of complex analysis in order to obtain $n$-point amplitudes in an efficient way.

There are in principle many ways of extending the real external momenta to complex quantities. We will deal explicitly with complex extensions of the momenta of two given particles, $i$ and $j$, being $\hat{p}_i$ and $\hat{p}_j$ the corresponding complex momenta. The particular complex extension described below is chosen so as to guarantee certain properties: (i) conservation of all the external momenta, (ii) that both $\hat{p}_i$ and $\hat{p}_j$ are null vectors, and (iii) that the complex poles associated with propagators are simple poles. These are satisfied by the complex shift
\begin{equation}
\hat{p}_i = p_i + z\eta,\qquad\qquad\hat{p}_j = p_j - z\eta,\label{eq:twoshiftdef}
\end{equation}
where $z\in\mathbb{C}$ and the vector $\eta$ has to satisfy certain conditions, namely $\eta\cdot p_i = \eta\cdot p_j =\eta^2 = 0$. Both conditions (i) and (ii) above guarantee that we can deal with scattering amplitudes of the shifted momenta in the same way as scattering amplitudes of real momenta; we will elaborate later in this section on the meaning of (iii). Hence under this shift, we extend the amplitude $A_n$ to a function $\hat{A}_n(z)$ with non-trivial dependence on the complex variable $z$.

The simplest scenario is that in which both particles $i$ and $j$ are massless particles (dealing with massive particles is explained in section \ref{sec:massive}). Then, we can write both orthogonality conditions as
\begin{equation}
\eta\cdot p_i =\frac{1}{2}\braket{\eta i}[\eta i] = 0,\qquad\qquad\eta\cdot p_j =\frac{1}{2}\braket{\eta j}[\eta j] = 0.\label{eq:nomassort}
\end{equation}
As two spinors are orthogonal if and only if they are proportional, there are two solutions to the equations above. Fixing the arbitrary proportionality constants to unity, one such solution to this system of equations is
 \begin{equation}
|\eta]=|j],\qquad \ket{\eta}=\ket{i}.\label{eq:ijshift0}
\end{equation}
The only other solution is physically equivalent under the exchange of particles $i$ and $j$ and multiplication of the complex variable $z$ by a $(-1)$ factor.

Decomposing all the elements in Eq. \eqref{eq:twoshiftdef} in terms of spinors, and using Eq. \eqref{eq:ijshift0}, the complex shift above is given in spinor-helicity notation by
\begin{equation}
|\hat{i}] = |i] + z|j],\qquad|\hat{j}] = |j],\qquad\ket{\hat{i}} = \ket{i},\qquad\ket{\hat{j}} = \ket{j} -z\ket{i}.\label{eq:ijshift}
\end{equation}
This is called a $[i,j\rangle$-shift. Note that for complex momenta, angle and square brackets are no longer related by complex conjugation, but rather are independent quantities. It is important to keep in mind this feature in order to properly understand the equation above. Indeed, it is the complex nature of the extension of external momenta which permits to write the shift \eqref{eq:twoshiftdef} in such a simple way in terms of spinor-helicity variables.

Now let us come back to the discussion about the shifted internal momenta $\hat{P}_{ab}$. If we choose the particles $a$ and $b$ to be the particles $i$ and $j$, from Eq. \eqref{eq:twoshiftdef} is easy to see that $\hat{P}_{ab}=P_{ab}$ and the internal momentum is not shifted. Hence let us consider $a=i$ and $b\neq j$. Then, using the equations above,
\begin{equation}
\hat{P}_{ib}^2 = (p_i+p_b+z\eta)^2 = P_{ib}^2 + 2z\,P_{ib}\cdot\eta = 0.
\end{equation}
As stated above, it is now straightforward to see that we can choose the complex part of the shifted momenta, or in other words the value of $z$, in order to ensure the on-shell nature of $\hat{P}_{ib}$. This value is given by
\begin{equation}
z_{ib} \equiv z|_{\hat{P}_{ib}^2=0} = \frac{-P_{ib}^2}{2P_{ib}\cdot\eta}.\label{eq:polesdef}
\end{equation}
Most importantly, the shifted $n$-point amplitude $\hat{A}_n(z)$ obtained shifting the momenta of particles $i$ and $j$ presents a simple pole at $z=z_{ib}$; namely, a singularity that behaves as $(z-z_{ib})^{-1}$:
\begin{equation}
\frac{1}{\hat{P}_{ib}^2}=\frac{1}{P_{ib}^2 + 2z\,P_{ib}\cdot\eta}=\frac{1}{2P_{ib}\cdot\eta}\frac{1}{z-z_{ib}}=-\frac{z_{ib}}{P_{ib}^2}\frac{1}{z-z_{ib}}.\label{eq:reshatP}
\end{equation}
The most relevant quantity associated to each pole of a given complex function is its residue, which is the finite value obtained when removing the singularity of the function by a suitable multiplicative factor. The residue of the complexified $n$-point amplitude can be evaluated using the standard definition of the residue of a simple pole:
\begin{equation}
\mbox{Res}_{z=z_{ib}}\hat{A}_n(z)=\lim_{z\rightarrow z_{ib}}(z-z_{ib})\hat{A}_n(z).\label{eq:resib}
\end{equation}
The rules that permit the evaluation of the residue in a simple way will be discussed in the next section.

In order to illustrate the relevance of poles for the evaluation of the physical $n$-point amplitude $A_n$, let us consider the slightly modified complex function $f(z)=\hat{A}_n(z)/z$ instead of $\hat{A}_n(z)$. For general external momenta the other poles of the function, of the form given in Eq. \eqref{eq:polesdef}, are away from $z=0$. Hence $\hat{A}_n(z)/z$ has a single pole at $z=0$, being the corresponding residue
\begin{equation}
\mbox{Res}_{z=0}f(z)=\lim_{z\rightarrow0}zf(z)=\hat{A}_n(0)=A_n.\label{eq:res0}
\end{equation}
This equation is not particularly useful by itself, as it just relates two different quantities, the $n$- point amplitude $A_n$ and its complex extension $\hat{A}_n(z)$, none of which are known in advance (but rather are the ones we are trying to obtain). However, as we discuss in the next section, Cauchy's residue theorem permit to combine Eqs. \eqref{eq:resib} and \eqref{eq:res0} in a way which, together with the simple rules that permit to evaluate the residue in Eq. \eqref{eq:resib}, leads to a useful expression for $A_n$ that can be applied in a variety of situations.

\subsubsection{Cauchy's residue theorem and subamplitudes \label{sec:cauchy}}

The poles of the complex function $f(z)=\hat{A}_n(z)/z$ described above can be related using Cauchy's residue theorem. This is one of the basics theorems in complex analysis which applies to meromorphic functions, namely complex functions which are differentiable (in the complex sense) everywhere but at its poles. Cauchy's residue theorem states that the integral of a complex function along a given closed curve that does not meet any of its poles equals the residues enclosed by the curve, up to a $2\pi i$ factor.

Let us consider a closed curve $\gamma$ which encloses all the poles of the function $f(z)$; for instance, a circle with radius $R\rightarrow\infty$. Then Cauchy's residue theorem implies that
\begin{equation}
B_n = \frac{1}{2\pi i}\oint_\gamma dz\,f(z) =\text{Res}_{z=0} \frac{\hat{A}_n(z)}{z}+ \sum_{z_{ib}} \text{Res}_{z=z_{ib}} \frac{\hat{A}_n(z)}{z},\label{eq:cauchy1}
\end{equation}
where we have defined the ``boundary term" $B_n$ as the integral over the curve $\gamma$ (the reason behind this notation will become clear in the next section).

From our previous discussion, the residue at $z=0$ yields back $A_n$. This permits to write the physical $n$-point amplitude (namely the quantity we want to evaluate) as
\begin{equation}
A_n = -\sum_{z_{ib}} \text{Res}_{z=z_{ib}} \frac{\hat{A}_n(z)}{z} + B_n.\label{eq:cauchy2}
\end{equation}

Admittedly, this equation may not seem specially useful in its current version. A first simplification stems from the fact that the boundary term vanishes in a large number of situations if the particles $i$ and $j$ being shifted are adequately chosen. We will discuss this in the next section; for the moment, let us put $B_n=0$ in the previous equation.

Then the $n$-point amplitude $A_n$ is determined entirely by the residues of $\hat{A}_n(z)/z$ at the $z=z_{ib}$ poles. The second simplification arises from the fact that each of these residues can be written as a product of on-shell lower-point amplitudes evaluated with complex momenta. In order to understand this feature, let us recall that the residue for a given pole $z=z_{ib}$ comes from a Feynman diagram with two particles $i$ and $b\neq j$ on one side of the propagator. The momentum flowing in the propagator, $\hat{P}_{ib}=\hat{p}_i+p_b$, becomes on-shell at the pole $z=z_{ib}$. When the momentum flowing in the propagator becomes on-shell, the amplitude represented by this complex Feynman diagram factorizes, so that
\begin{equation}
\mbox{Res}_{z=z_{ib}}\frac{\hat{A}_n(z)}{z} = -i A_{L}(z_{ib})\frac{1}{P_{ib}^2}A_{R}(z_{ib}).\label{eq:polefact}
\end{equation}
Here $A_L(z_{ib})$ is the on-shell amplitude for the particles $i$, $b\neq j$ and the one in the propagator, and $A_R(z_{ib})$ the on-shell amplitude for the particle in the propagator and the remaining particles in the $n$-point amplitude.

It is not our aim to prove this factorization property here, but we can illustrate that it is reasonable using a fairly general example. Let us consider an $n$-particle interaction mediated by a virtual fermion with spin $1/2$, with arbitrary external particles. Then any non-zero complex amplitude associated with a given shifted Feynman diagram would have the structure
\begin{equation}
g(z)=-i\frac{\langle \hat{X}|\hat{P}_{ib}|\hat{Y}]}{\hat{P}_{ib}^2}.
\end{equation}
Here $|\hat{X}\rangle$ and $|\hat{Y}]$ depend on the particular kinds of external particles being considered (note that these spinors are shifted). Taking into account that the pole in $z=z_{ib}$ appears due to the denominator, due to Eq. \eqref{eq:reshatP}, it is straightforward to check that the residue of the quantity above divided by $z$ is given by
\begin{equation}
\mbox{Res}_{z=z_{ib}}\frac{g(z)}{z}=\frac{i}{P_{ib}^2}\langle \hat{X}|\hat{P}_{ib}|\hat{Y}]=-i\langle \hat{X}\hat{P}_{ib}\rangle\frac{1}{P_{ib}^2}[\hat{P}_{ib}\hat{Y}].\label{eq:factexmp}
\end{equation}
In the last equation we have exploited the property that, when $\hat{P}_{ib}$ is on-shell, it can be written in terms of spinors as $\hat{P}_{ib}=-|\hat{P}_{ib}\rangle[\hat{P}_{ib}|-|\hat{P}_{ib}]\langle\hat{P}_{ib}|$. But Eq. \eqref{eq:factexmp} displays the structure of Eq. \eqref{eq:polefact}: $\langle\hat{X}\hat{P}_{ib}\rangle$ is the on-shell amplitude for the particles to the left of the propagator and the spin-1/2 fermion, while $[\hat{P}_{ib}\hat{Y}]$ is the corresponding quantity for the particles on the other side of the propagator. These on-shell amplitudes necessarily involve less than $n$ particles, and are evaluated in the particular values of the complex momenta determined by the condition $z=z_{ib}$.

Remarkably, this factorization property holds in general. Hence we can write each residue in the left-hand side of Eq. \eqref{eq:cauchy2} as in Eq. \eqref{eq:polefact}, so that the $A_n$ amplitude is given by
\begin{equation}
A_n = i\sum_{z_{ib}}\sum_{h}A_L(z_{ib})\frac{1}{P_{ib}^2}A_R(z_{ib}).\label{eq:recrel}
\end{equation}
This equation contains an additional sum over $h$, the index corresponding to the helicity of the internal particle. In the example given above with an internal spin-1/2 fermion this sum was implicit, through the use of the relation $\hat{P}_{ib}=-|\hat{P}_{ib}\rangle[\hat{P}_{ib}|-|\hat{P}_{ib}]\langle\hat{P}_{ib}|$, though one of the contributions were identically zero.

Eq. \eqref{eq:recrel} condenses the content of the celebrated BCFW recursion relations \cite{Britto:2004ap,Britto:2005fq,Feng:2011np}. In general, this equation implies that a given $n$-point amplitude can be written as a sum of products of lower-point amplitudes, where the sum has to be taken on arrangements of external particles that guarantee that the internal momenta are shifted, as well as on the helicity of internal particles. Knowing what we do about little group scaling and 3-point kinematics, we can now break down any complicated amplitude of $n$ particles into products of 3-point amplitudes which are themselves easily calculated. This will be thoroughly illustrated in examples associated with particular processes of astrophysical significance.

\subsubsection{Showing shift-validity \label{sec:valid}}

The simplicity of the recursion relations in Eq. \eqref{eq:recrel} above rests on the assumption that the boundary term vanishes, $B_n=0$. Using the definition of this quantity in Eq. \eqref{eq:cauchy1}, namely
\begin{equation}
B_n = \frac{1}{2\pi i}\oint_\gamma dz\,f(z),
\end{equation}
and considering a contour that goes to infinity, this is equivalent to $\lim_{|z|\rightarrow\infty}zf(z)=0$. In terms of the complex amplitude $\hat{A}_n(z)$, this condition translates into
\begin{equation}
\lim_{|z|\rightarrow\infty}\hat{A}(z)=0.\label{eq:validdef}
\end{equation}

Hence in order to ensure that usage of the recursion relations is justified, it is necessary to ensure that the complex extension of the $n$-point amplitude that we want to evaluate decays to zero in the limit of complex infinity.  
Not every imaginable shift choosing arbitrary particles $i$ and $j$ would verify this constraint; the combinations of shifted particles that satisfy this condition are known as ``valid'' shifts.  

There is no general rule to follow in order to show that a given shift is valid, but different situations require different approaches. The general strategy is to show that Eq. \eqref{eq:validdef} is satisfied by looking for a bound on the leading behavior of $\hat{A}(z)$ at complex infinity, namely $|\hat{A}(z)|\leq k |z|^{-\alpha}$ for some real constant $k$ and positive real exponent $\alpha>0$. 

The simplest bound that can be obtained follows from the behavior of all individual Feynman diagrams that contribute to a given amplitude. The evaluation of a given $n$-point amplitude using on-shell recursion relations represents a more effective determination of all the\linebreak contributions to a given process coming from different Feynman diagrams. However, at the end of the day the results of both evaluations have to be the same. In particular, the leading behavior with $z$ of complex $n$-point amplitude $\hat{A}(z)$ has to be the same as the leading behavior of the sum of all individual Feynman diagrams contributing to the complex amplitude. Hence an upper bound to the asymptotic behavior of $\hat{A}(z)$ at complex infinity can be obtained as the leading contribution from the dominant individual Feynman diagram at large $|z|$. This is an upper bound due to the fact that when summing all the contributions coming from the different Feynman diagrams, cancellations of the apparent leading term in $z$ can take place.

What makes this bound useful is that it is straightforward to compute. Feynman diagrams are a product of different elements: external legs, interaction vertices, and propagators. It is not difficult to obtain the leading behavior with $z$ of each of these elements, and then take their product in order to obtain the leading behavior with $z$ for a given Feynman diagram. For instance, let us consider an $[i,j\rangle$-shift where the helicities of the shifted particles are $(h_i,h_j) = (-1,+1)$. Due to Eq. \eqref{eq:reshatP}, the internal propagator $1/\hat{P}^2_{ib}$ behaves as $z^{-1}$. Individual polarization vectors are given in Eq. \eqref{eq:polvec} and, for the chosen shift, also lead to a $z^{-1}$ dependence; for instance,
\begin{equation}
\epsilon^\mu_+(\hat{p}_j;q) = -\frac{\bra{q}\gamma^\mu|j]}{\sqrt{2}\braket{q\hat{j}}}.
\end{equation}
Note that, for the same helicities of external particles, the alternative $[j,i\rangle$-shift leads to a leading behavior linear in $z$ instead. Hence it is important to choose properly the shift in order to use this method to show its validity. Lastly, the behavior of interaction vertices depends on the particular theory being considered. With all these ingredients, it is possible to extract the leading behavior with $z$ of individual Feynman diagrams, hence obtaining a bound to the asymptotic behavior of a given (complex) $n$-point amplitude $\hat{A}_n(z)$. This method works in a large number of situations, provided a wise choice of shift is made. This will be illustrated through the examples presented in the main body of the paper. However, it may occur that this bound is not tight enough to show that at least one shift for a given $n$-point amplitude is valid. In these situations, the only systematic way to proceed is to consider shifts of more than two particles, which may improve the leading behavior with $z$ of individual Feynman diagrams \cite{Cohen:2010mi,Feng:2011np}. If this does not work there is no general rule to apply, though it is customary to refine the bounds coming from the leading behavior of individual Feynman diagrams by checking for potential cancellations of the apparent leading terms in $z$.

\subsubsection{BCFW with massive particles \label{sec:massive}}

The BCFW relations for two shifted particles can be easily extended to include a massive particle (shifting two massive particles forbids solving explicitly for the shift in terms of spinor-helicity variables \cite{Badger:2005zh,Badger:2005jv}). As in the massless case, we introduce a null vector $\eta^\mu$ and consider the shift of the external momenta
\begin{equation}
\hat{p}_i=p_i+z\eta,\qquad\qquad \hat{p}_j=p_j-z\eta.\label{eq:shift1}
\end{equation}
However, let us now assume that the particle $j$ is massive. This makes the determination of $\eta^\mu$ slightly different, as the orthogonality relation $\eta\cdot p_j$ now takes a different form. Since $\eta^\mu$ is still null, we can decompose $\eta^{\dot{a}b}$ as
\begin{equation}
\eta^{\dot{a}b}=-|\eta\rangle^{\dot{a}}[\eta|^b.
\end{equation}
This null vector has to be orthogonal to $p_i$ and $p_j$. The first of these conditions is the same as in the purely massless case, namely the first condition in Eq. \eqref{eq:nomassort}:
\begin{equation}
\eta\cdot p_i=\frac{1}{2}\langle\eta i\rangle[\eta i]=0.
\end{equation}
This requires either $|\eta\rangle^{\dot{a}}=|i\rangle^{\dot{a}}$ or $[\eta|^b=[i|^b$. Let us consider explictitly the first solution, that corresponds to the analogue of the $[i,j\rangle$-shift (the discussion in the alternative case is completely parallel).

On the other hand, the second orthogonality condition is now written as
\begin{equation}
\eta\cdot p_j=\frac{1}{2}[\eta|^b (p_j)_{b\dot{a}}|\eta\rangle^{\dot{a}}=\frac{1}{2}[\eta|^b (p_j)_{b\dot{a}}|i\rangle^{\dot{a}}=0.\label{eq:jmassort}
\end{equation}
If the particle $j$ was massless, we could use the decomposition of $p_j$ in terms of spinors to obtain the second orthogonality condition in Eq. \eqref{eq:nomassort}. In any case, for $j$ massive it is still possible to solve explicitly Eq. \eqref{eq:jmassort} as
\begin{equation}
[\eta|^b=\epsilon^{bc}(p_j)_{c\dot{a}}|i\rangle^{\dot{a}}.\label{eq:1bisp}
\end{equation}
In summary, the equivalent of Eq. \eqref{eq:ijshift} is now given by
\begin{align}
&|\hat{i}]=|i]+zp_j|i\rangle,\qquad |\hat{i}\rangle=|i\rangle,\qquad \hat{p}_j^{\dot{a}b}=p_j^{\dot{a}b}-z|i\rangle^{\dot{a}}\epsilon^{bc}(p_j)_{c\dot{b}}|i\rangle^{\dot{b}}.\label{eq:shiftmassive1}
\end{align}
The rest of the discussion is parallel to the massless case, but taking into account that for internal particles with mass $m_{ib}$, poles arise for the values $z=z_{ib}$ that make $\hat{P}^2_{ib}=-m_{ib}^2$. The equivalent of Eq. \eqref{eq:recrel} is now given by
\begin{equation}
A_n = i\sum_{z_{ib}}\sum_{h}A_L(z_{ib})\frac{1}{P_{ib}^2+m_{ib}^2}A_R(z_{ib}),\label{eq:recrelmass}
\end{equation}
where $m_{ib}$ is the mass of the internal particle for the partition in which the particles labelled by $i$ and $b$ are on the same side of the propagator.

\section{Light bending and gravitational wave scattering from BCFW}

\subsection{Light bending}

Having spent the previous section building the technology of BCFW relations, let's now put it to use by revisiting, once more, the gravitational lensing of light by a massive body. As explained in section \ref{sec:lbqft}, gravitational lensing of light can be evaluated using the techniques of quantum field theory. In this section we show how this evaluation is greatly simplified when using on-shell techniques.

\subsubsection{Evaluation of the scattering amplitude}

We consider the gravitational force between a massive, spinless object (such as a non-rotating star) and a massless photon. Diagrammatically, this looks like:\bigskip
\begin{equation}\label{fdiags1}
\begin{gathered}
\begin{fmfgraph*}(120,80)
\fmfleft{i1,i2}
\fmfright{o1,o2}
\fmf{photon,label=$1^{\mp1}$,label.side=left}{i1,v1}
\fmf{plain,label=$4^0$,label.side=left}{i2,v1}
\fmf{photon,label=$2^{\pm1}$,label.side=left}{v1,o1}
\fmf{plain,label=$3^0$,label.side=left}{v1,o2}
\fmfblob{.30w}{v1}
\end{fmfgraph*}
\end{gathered}
~~~=~~~
\begin{gathered}
	\begin{fmfgraph*}(120,80)
	\fmfleft{i1,i2}
	\fmfright{o1,o2}
	\fmf{photon,label=$1^{\mp1}$,label.side=left}{i1,v1}
	\fmf{fermion,label=$4^0$,label.side=left}{i2,v2}
	\fmf{dbl_wiggly}{v1,v2}
	\fmf{photon,label=$2^{\pm1}$,label.side=left}{v1,o1}
	\fmf{fermion,label=$3^0$,label.side=left}{v2,o2}
	\fmflabel{$_{+2}$}{v1}
	\fmflabel{$_{-2}$}{v2}
	\end{fmfgraph*}
\end{gathered}
~~~+~~~
\begin{gathered}
\begin{fmfgraph*}(120,80)
\fmfleft{i1,i2}
\fmfright{o1,o2}
\fmf{photon,label=$1^{\mp1}$,label.side=left}{i1,v1}
\fmf{fermion,label=$4^0$,label.side=left}{i2,v2}
\fmf{dbl_wiggly}{v1,v2}
\fmf{photon,label=$2^{\pm1}$,label.side=left}{v1,o1}
\fmf{fermion,label=$3^0$,label.side=left}{v2,o2}
	\fmflabel{$_{-2}$}{v1}
	\fmflabel{$_{+2}$}{v2}
\end{fmfgraph*}
\end{gathered}
\end{equation}

Notice that we are not explicitly including symmetrization with respect to identical particles. Following our conventions, we imagine this as particles 1 and 4 incoming, exchanging a graviton and then outgoing with momentum 2 and 3. Particles 3 and 4 are massive, with on-shell condition $p_3^2 = p_4^2 = -m^2$. The action corresponding to such an interaction is the Einstein-Maxwell-Scalar action previously introduced in Eq. \eqref{eq:EMSact}.

Following our discussion on the BCFW relations, the goal in this section is to evaluate the 4-point amplitude depicted in Fig. \ref{fdiags1} from the knowledge of the relevant 3-point (sub-)amplitudes. Let us therefore start by calculating the two 3-point amplitudes associated with this diagram. The first one is given by:
\begin{equation}
\underset{(1)}{
	\begin{fmfgraph*}(150,80)
	\fmfleft{i1,i2}
	\fmfright{o1}
	\fmf{photon,label=$1^{\mp1}$}{i1,v1}
	\fmf{photon,label=$2^{\pm1}$}{i2,v1}
	\fmf{dbl_wiggly, label=$P_{12}^{\pm2}$}{v1,o1}
	\end{fmfgraph*}
}
\end{equation}
Recall that $P_{12}=p_1+p_2$. Since this first diagram involves only massless particles, we can use little group scaling to calculate the associated 3-point amplitude. 

From all the possible helicity choices, we can now show that in order to evaluate the light-bending angle it is only necessary to consider these in which $h_1$ and $h_2$ are different. Amplitudes such that $h_1=h_2$ describe a change of helicity of the photon, due to its gravitational interaction with the massive scalar field. This process is not allowed. The simplest way to show this is using dimensional analysis. By dimensional analysis, we can see that the Ricci scalar $R$ is of mass dimension $2$, since it contains two derivatives of the dimensionless field $h_{\mu\nu}$. The coupling constant $\kappa$ must then have mass dimension $-1$, so that the total mass dimension of $2R/\kappa^2$ is 4 as required (to ensure that the action is dimensionless). The angle and square brackets have mass dimension $1$, since each corresponds to some (possibly complex) momentum. As explained at the end of section \ref{sec:lgs}, we require that any amplitude involving $n$ external legs hass mass dimension $4-n$ in 4 dimensions, thus we require that $[A] = 1$.

Let us consider for instance the helicity configuration $(h_1,h_2,h_3)=(-1,-1,+2)$. Little group scaling tells us that the amplitude $A_3(1^{-1}2^{-2}P_{12}^{+2})$ is either given by
\begin{equation}
-\frac{\kappa}{2}\frac{\braket{12}^4}{\braket{1P_{12}}^2\braket{2P_{12}}^2}
\end{equation}
or
\begin{equation}
-\frac{\kappa}{2}\frac{[1P_{12}]^2[2P_{12}]^2}{[12]^4}.
\end{equation}
It is straightforward to see that any of these expressions have the correct dimensionality. This means that this helicity choice does not contribute to the scattering amplitude. A similar argument applies to the choices $(h_1,h_2,h_3)=(+1,+1,-2)$, $(+1,+1,+2)$ and $(-1,-1,-2)$.

These leave us with only two possibilities. For instance, for the choice $(h_1,h_2,h_3)=(+1,-1,-2)$, one has:
\begin{equation}
A_3(1^{+1}2^{-1}P_{12}^{-2}) = -\frac{\kappa}{2}\frac{\langle2P_{12}\rangle^4}{\langle12\rangle^2}.\label{eq:p3ptb}
\end{equation}
As discussed in section \ref{sec:lgs}, arguments based in little group scaling leave freedom to consider either the square bracket version or the angle bracket version of a given 3-point amplitude. However, here is where dimensional analysis enters into play. The amplitude in Eq. \eqref{eq:p3pt} satisfies this requirement, as it has dimension 1. On the other hand, the angle bracket version of $A(1^{-1}2^{+1}P_{12}^{+2})$ as determined by little group scaling would be proportional to $\kappa\langle12\rangle^2\langle2P_{12}\rangle^4$, which clearly does not display the correct dimensions, and therefore can be discarded.

Making parallel arguments, the remaining choice of helicities, $(h_1,h_2,h_3)=(-1,+1,+2)$, leads to
\begin{equation}
A_3(1^{-1}2^{+1}P_{12}^{+2}) = -\frac{\kappa}{2}\frac{[2P_{12}]^4}{[12]^2}.\label{eq:p3pt}
\end{equation}

The next 3-point amplitude that we have to evaluate is diagrammatically represented by
\begin{equation}
\underset{(2)}{
	\begin{fmfgraph*}(150,80)
	\fmfleft{i1,i2}
	\fmfright{o1}
	\fmf{plain,label=$3^0$}{i1,v1}
	\fmf{plain,label=$4^0$}{i2,v1}
	\fmf{dbl_wiggly, label=$-P_{12}^\mp$}{v1,o1}
	\end{fmfgraph*}
}
\end{equation}
Note that we are exploiting the fact that, due to momentum conservation, $P_{12}=-p_3-p_4$. To evaluate this diagram we cannot resort to little group scaling, as two of the legs are massive. Hence we have to use the relevant vertex, obtained as part of the Feynman rules of the theory:
\begin{equation}
V^{\mu\nu}(3^04^0) = \frac{-i\kappa}{2}\left[p^\mu_3p^\nu_4 + p^\nu_3p^\mu_4 - \eta^{\mu\nu}(p_3\cdot p_4 -m^2)\right].
\end{equation}
We dot this with our chosen polarization vectors, in this case $\epsilon^\pm(-P_{12})_\mu\epsilon^\pm(-P_{12})_\nu$, to find the two possible 3-point amplitudes:
\begin{equation}
A_3((-P_{12})^{-2}3^04^0) = \frac{-i\kappa}{2}\left[2(p_3\cdot\epsilon^-)(p_4\cdot\epsilon^-)\right] = \frac{-i\kappa\bra{P_{12}}p_3|q]\bra{P_{12}}p_4|q]}{2[qP_{12}]^2}\label{eq:zz1}
\end{equation}
and
\begin{equation}
A_3((-P_{12})^{+2}3^04^0) = \frac{-i\kappa}{2}\left[2(p_3\cdot\epsilon^+)(p_4\cdot\epsilon^+)\right] = \frac{-i\kappa\bra{q}p_3|P_{12}]\bra{q}p_4|P_{12}]}{2\braket{qP_{12}}^2}\label{eq:zz2}
\end{equation}
In these equations, we have used $|-p\rangle=-|p\rangle$ and $|-p]=+|p]$ (see \cite{Elvang:2013cua} for instance).

It is a good moment to introduce a particular manipulation involving spinors and momenta that will be used often below (most of the times, implicitly). Let us take for instance the piece $\langle P_{12}|p_4|q]$ in Eq. \eqref{eq:zz1}. Due to momentum conservation, $p_4=P_{34}-p_3=-P_{12}-p_3$, so that
\begin{equation}
\langle P_{12}|p_4|q]=-\langle P_{12}|P_{12}|q]-\langle P_{12}|p_3|q]=-\langle P_{12}|p_3|q].
\end{equation}
The term $\langle P_{12}|P_{12}|q]$ is identically zero using the relevant form of the Weyl equation in Eq. \eqref{eq:weyleq}, as $\langle P_{12}|P_{12}=0$. This permits to simplify the form of the 3-point amplitude \eqref{eq:zz1}. This is one of the manipulations that permits huge simplifications of scattering amplitudes.

Hence we have determined the relevant 3-point amplitudes to be used. In the following let us focus in the case $(h_1,h_2)=(+1,-1)$. The steps to be followed for the complementary case $(h_1,h_2)=(-1,+1)$ are identical but, most importantly, this other case can be obtained straightforwardly by exchanging the particles $1$ and $2$ in the final expression of $A_4(1^{+1}2^{-1}3^04^0)$. The next step is choosing the momentum that we wish to make complex, which we will take to be the adjacent momenta $2$ and $3$, that is, choosing one massive and one massless. In principle, the choice of shift is dictated by the requirement of being valid (in the sense defined in section \ref{sec:valid}). However, in this particular example it is not possible to show that any of the possible two-particle shifts is valid. For instance, for the choice we are going to follow below, namely a $[2,3\rangle$-shift, the leading behavior of individual Feynman diagrams is as follows. Each interaction vertex contributes a factor of $z$, since each numerator depends on at least one shifted momenta. The propagator contributes a factor of $1/z$, and the shifted polarization vector $\epsilon^\mu_-(p_2)$ will contribute a factor of $1/z$ (recall Eq. \eqref{eq:polvec}). This means that in total, our $z$-dependence is actually $z^0$, which means in principle there could be a boundary contribution to this process. However, we can justify using this shift by implementing an ``auxiliary'' shift, as described in \cite{Badger:2005zh,Hall:2008xn}. We will not go through the details here as it involves introducing additional technical machinery, but we have checked that using a three-particle shift as an auxiliary shift permits to show that the shift we are using is valid. In the example of the next section, we will show explicitly the validity of the corresponding two-particle shift.
 
To apply the $[2,3\rangle$-shift to this case we just have to particularize the general discussion in section \ref{sec:massive} to $i=2$ and $j=3$ in the present discussion. Eq. \eqref{eq:shiftmassive1} leads to
\begin{align}
&\hat{p}_2^{a\dot{b}} = \ket{2}^a[\hat{2}|^{\dot{b}} = \ket{2}^a[2|^{\dot{b}} + z\ket{2}^a(\bra{2}\slashed{p}_3)^{\dot{b}},\nonumber\\
&\hat{p}_3^{a\dot{b}} = p_3^{a\dot{b}} -z\ket{2}^a(\bra{2}\slashed{p}_3)^{\dot{b}}.
\end{align}
For simplicity, we will mostly write these as
\begin{equation}
\hat{p}_2 = \ket{2}[2| + z\ket{2}[\eta|,\qquad\hat{p}_3 = p_3 - z\ket{2}[\eta|,
\end{equation}
where
\begin{equation}
|\eta]=p_3|2\rangle.
\end{equation}
For this choice of shifted particles, there is only one possible arrangements of the particles: $\hat{2}$ and 4 on one side (e.g., left) of the shifted propagator, and $1$ and $\hat{3}$ on the other side (that is, on the right). The application of the recursion relations \eqref{eq:recrel} lead therefore to
\begin{align}
A_4(1^{+1}2^{-1}3^04^0)&=\frac{i}{P_{12}^2}A_3(1^{+1}\hat{2}^{-1}\hat{P}_{12}^{-2})A_3((-\hat{P}_{12})^{+2}\hat{3}^04^0)\nonumber\\
&+\frac{i}{P_{12}^2}A_3(1^{+1}\hat{2}^{-1}\hat{P}_{12}^{+2})A_3((-\hat{P}_{12})^{-2}\hat{3}^04^0),\label{eq:nn1}
\end{align}
where shifted momenta are evaluated on $z=z_{12}$ that guarantees that the shifted momentum in the propagator, $\hat{P}_{12}=p_1+\hat{p}_2$, is on-shell. Taking into account Eq. \eqref{eq:viprel}, this means that
\begin{equation}
\hat{P}_{12}^2 =2p_1\cdot\hat{p}_2= \braket{12}[1\hat{2}] = \braket{12}\left([12] + z_{12}[1|p_3|2\rangle\right) = 0
\end{equation}
or, equivalently,
\begin{equation}
[1\hat{2}]=[12] + z_{12}[1|p_3|2\rangle=0,\label{eq:12hat0}
\end{equation}
which permits to obtain
\begin{equation}
z_{12}= -\frac{[12]}{[1|p_3|2\rangle}.\label{eq:nn2}
\end{equation}
Eqs. \eqref{eq:nn1} and \eqref{eq:nn2}, together with the 3-point amplitudes evaluated just above, are all we need to obtain the 4-point amplitude. For this, the only necessary step is writing the 3-point amplitudes in terms of the shifted momenta. Let us start with the first line of Eq. \eqref{eq:nn1}. This term contains the following 3-point amplitude:
\begin{equation}
A_3(1^{+1}\hat{2}^{-1}\hat{P}_{12}^{+2}) = -\frac{\kappa}{2}\frac{[1\hat{P}_{12}]^4}{[1\hat{2}]^2} = -\frac{\kappa}{2}\frac{[1\hat{2}]^2\braket{12}^4}{\braket{1\hat{P}_{12}}^4} = 0.
\end{equation}
To write the second identity we have multiplied by $\langle 1\hat{P}_{12}\rangle^4/\langle 1\hat{P}_{12}\rangle^4$. That the quantity above vanishes follows then from Eq. \eqref{eq:12hat0}. This means that the first term in Eq. \eqref{eq:nn1} does not contribute to the amplitude.

Let us consider the only remaining contribution, namely the second line in Eq. \eqref{eq:nn1}. The two 3-point amplitudes involved are
\begin{align}
A(1^{+1}\hat{2}^{-1}\hat{P}_{12}^{-2}) = -\frac{\kappa}{2}\frac{\braket{2\hat{P}_{12}}^4}{\braket{12}^2}
\end{align}
and
\begin{equation}
A((-\hat{P}_{12})^{+2}\hat{3}^04^0)= \frac{-i\kappa}{2}\frac{\bra{q}\hat{p}_3|\hat{P}_{12}]\bra{q}p_4|\hat{P}_{12}]}{\braket{q\hat{P}_{12}}^2}=\frac{i\kappa}{2}\frac{\bra{q}p_4|\hat{P}_{12}]^2}{\braket{q\hat{P}_{12}}^2},
\end{equation}
where we have used $\hat{p}_3=-p_4-\hat{P}_{12}$.

Hence the full 4-point amplitude is given by
\begin{align}
A_4(1^{+1}2^{-1}3^04^0) &= A(1^{+1}\hat{2}^{-1}\hat{P}_{12}^{-2})\frac{i}{P_{12}^2}A((-\hat{P}_{12})^{+2}\hat{3}^04^0)\nonumber\\
&= \left(-\frac{\kappa}{2}\frac{\braket{2\hat{P}_{12}}^4}{\braket{12}^2}\right)\frac{i}{P_{12}^2}\left(\frac{i\kappa\bra{q}p_4|\hat{P}_{12}]^2}{2\braket{q\hat{P}_{12}}^2}\right)\nonumber\\
&= \frac{\kappa^2}{4}\frac{1}{P_{12}^2}\frac{\bra{q}p_4|\hat{P}_{12}]^2\braket{2\hat{P}_{12}}^4}{\braket{12}^2\langle q\hat{P}_{12}\rangle^2}.
\end{align}
This expression can be simplified using
\begin{equation}
\bra{q}p_4|\hat{P}_{12}]\braket{P_{12}2}=-\bra{q}p_4|1]\braket{12},
\end{equation}
leading to the following closed form of the amplitude:
\begin{align}
A_4(1^{+1}2^{-1}3^04^0)=\frac{\kappa^2}{4}\frac{\bra{q}p_4|1]^2}{P_{12}^2}\frac{\braket{2\hat{P}_{12}}^2}{\langle q\hat{P}_{12}\rangle^2}.\label{eq:nn3}
\end{align}
This expression is written in terms of the arbitrary reference bispinor $\langle q|$. To evaluate the cross-section, it is necessary to obtain the modulus (in the complex sense) of Eq. \eqref{eq:nn3}. It can be show explicitly that, in this step, the dependence in this arbitrary reference spinor drops off, which means that physical results do not depende on this choice (physical results are gauge invariance). Hence, for the sake of simplicity, we can simplify the expression above if we take $\langle q|=\langle2|$, which leads to the simpler expression
\begin{align}
A_4(1^{+1}2^{-1}3^04^0)=\frac{\kappa^2}{4}\frac{\bra{2}p_4|1]^2}{P_{12}^2}.\label{eq:nn4}
\end{align}
%

\subsubsection{Evaluation of the cross-section}

Following the same steps as in section \eqref{sec:lbqft}, Eq. \eqref{eq:nn4} is not enough to extract the physics of scattering events. It is necessary to compute the modulus of this (generally complex) quantity, in order to obtain the cross-section. Also in this procedure, spinor-helicity variables will be transformed to momentum (or Mandelstam) variables in order to make easier the interpretation of the result. This is what we do in this section.

First of all, using the rule $\bra{2}p_4|1]^\dagger = \bra{1}p_4|2]$, the modulus of $A_4(1^{+1}2^{-1}3^04^0)$ can be written as
\begin{equation}
|A_4(1^{+1}2^{-1}3^04^0)|^2 = \frac{\kappa^4}{16}\frac{(\bra{2}p_4|1]\bra{1}p_4|2])^2}{P^4_{12}}.\label{eq:mm1}
\end{equation}
For the next step, it is instructive to make the spinor indices explicit, in order to show that the spinors in the numerator constitute the trace of four momentum vectors:
\begin{align}
&(\bra{2}p_4|1]\bra{1}p_4|2])^2\nonumber\\
&=(\bra{2}_{\dot{a}}(p_4)^{\dot{a}b}|1]_b\bra{1}_{\dot{c}}(p_4)^{\dot{c}d}|2]_d)^2\nonumber\\
&=((p_4)^{\dot{a}b}|1]_b\bra{1}_{\dot{c}}(p_4)^{\dot{c}d}|2]_d\bra{2}_{\dot{a}})^2\nonumber\\
&=((p_4)^{\dot{a}b}(p_1)_{b\dot{c}}(p_4)^{\dot{c}d}(p_2)_{d\dot{a}})^2.
\end{align}
Hence we recognise that the square object in the numerator of Eq. \eqref{eq:mm1} is none other than a trace. Hence the modulus of the scattering amplitude can be written purely in terms of momentum vectors as
\begin{equation}
|A_4(1^{+1}2^{-1}3^04^0)|^2 = \frac{\kappa^4}{16}\frac{\left(\Trm{p_4p_1p_4p_2}\right)^2}{P^4_{12}}.
\end{equation}
We have recognised this as the negative trace due to the ordering of indices, recalling that in the equation above, momenta are bispinors given by $p^{\dot{a}b} \equiv p_\mu({\bar{\sigma}}^{\mu})^{\dot{a}b}$. Dealing with this kind of expression is facilitated by the properties of the trace of Pauli matrices. These properties are well-known and can be found in many sources \cite{Peskin:1995ev,Zee:2003mt}. For instance, in our case we can use
\begin{align}\label{identities}
\Trm{\overline{\sigma}^\mu\sigma^\nu\overline{\sigma}^\rho\sigma^\lambda} &= 2(\eta^{\mu\nu}\eta^{\rho\lambda} - \eta^{\mu\rho}\eta^{\nu\lambda} + \eta^{\mu\lambda}\eta^{\rho\nu} - i\epsilon^{\mu\nu\rho\lambda} ).
\end{align}
When evaluating traces of slashed momenta, this becomes:
\begin{align}\label{key}
&\Trpm{p_4p_1p_4p_2} \nonumber\\
&= 2\left[(p_4\cdot p_1)(p_4\cdot p_2) - (p_4\cdot p_4)(p_1\cdot p_2) + (p_4\cdot p_2)(p_1\cdot p_4) \pm i\epsilon_{\mu\nu\rho\lambda}(p_4)^\mu (p_1)^\nu (p_4)^\rho (p_2)^\lambda\right].
\end{align}
Since two of the momentum vectors are the same within the trace, the antisymmetric component of this is zero, namely $\epsilon_{\mu\nu\rho\lambda}(p_4)^\mu (p_1)^\nu (p_4)^\rho (p_2)^\lambda=0$, and we are left with the simpler result
\begin{align}\label{key}
|A_4(1^{+1}2^{-1}3^04^0)|^2 &= \frac{\kappa^4}{8}\frac{\left[(p_4\cdot p_1)(p_4\cdot p_2) - (p_4\cdot p_4)(p_1\cdot p_2) + (p_4\cdot p_2)(p_1\cdot p_4)\right]^2}{P^4_{12}}\nonumber\\
 &= \frac{\kappa^4}{8}\frac{\left[2(p_1\cdot p_4)(p_2\cdot p_4) + m^2(p_1\cdot p_2)\right]^2}{(p_1 + p_2)^4}.
\end{align}
In the second equation we have used the on-shell condition $p_4=-m^2$ to simplify one of the three terms, and notice that the remaining two ones are indeed the same.

We can now write this in terms of Mandelstam invariants, defined in Eq. \eqref{eq:Mandelstam}, and given explicitly by
\begin{align}
&s_{12}= -(p_1 + p_2)^2 = -2p_1\cdot p_2,\nonumber\\
&s_{13} = -(p_1 + p_3)^2 = m^2 - 2p_1\cdot p_3,\nonumber\\
&s_{14} = -(p_1 + p_4)^2 = m^2 - 2p_1\cdot p_4.
\end{align}
The second identities above can be used by just using the on-shell conditions on the different momenta (both massless and massive). These invariants satisfy
\begin{equation}
s_{12} + s_{13} + s_{14} = -2m^2.\label{eq:manduse}
\end{equation}
In terms of Mandelstam variables, the modulus of the amplitude reads
\begin{equation}
|A_4(1^{+1}2^{-1}3^04^0)|^2 = \frac{\kappa^4}{16}\frac{\left[s_{13}s_{14} + m^2(s_{12}+s_{13}+s_{14}) + m^4\right]^2}{s_{12}^2} = \frac{\kappa^4}{16}\frac{\left(s_{13}s_{14} - m^4\right)^2}{s_{12}^2}.
\end{equation}
We have used Eq. \eqref{eq:manduse} to write the second identity above. This last expression of the modulus of the amplitude is manifestly invariant under the exchange of particles $3$ and $4$.

We are now interested in calculating the cross-section of this interaction. We choose again the centre of mass frame, in which the cross-section is given by the formula:
\begin{equation}\label{crosssec}
\frac{d\sigma}{d\Omega} = \frac{1}{64\pi^2 s_{14}}|A_4(1^{+1}2^{-1}3^04^0)|^2.
\end{equation} 
In the low energy limit, we imagine that the photon's energy as small compared with the scalar mass $m$. The approximations that can be considered in this limit and in this frame where explained in the bullet list in between Eqs. \eqref{eq:befcm} and \eqref{eq:aftcm}. This leads to the following simplified expressions of the Mandelstam invariants:
\begin{align}
&s_{12}\simeq  \vec{P}^2 = 4E^2 \sin^2(\theta/2),\nonumber\\
&s_{13} \simeq  m^2 - 2mE - 4E^2 \sin^2(\theta/2),\nonumber\\
&s_{14} \simeq (m+E)^2 \simeq m^2 + 2mE.
\end{align}
Furthermore, if we take the scattering angle to be small, then $\sin(\theta/2) \simeq \theta/2$.

Under all the simplifications, we can write the cross-section in this limit as
\begin{align}
\frac{d\sigma}{d\Omega} = \frac{1}{64\pi^2 s_{14}}|A_4(1^{+1}2^{-1}3^04^0)|^2= &\frac{\kappa^4}{16}\frac{\left[4m^2E^2+ 4m^2E^2\sin^2(\theta/2)\right]^2}{1024\pi^2m^2E^4\sin^4(\theta/2)}\\
&\simeq \frac{\kappa^4 m^2}{1024 \pi^2}\left(\frac{\sin^2(\theta/2) + 1}{\sin^2(\theta/2)}\right)^2\simeq \frac{\kappa^4m^2}{64 \pi^2\theta^4},
\end{align}
where we have used that $m + 2E \simeq m$ and $\sin^2(\theta/2) \ll 1$. We can now recall that $\kappa^2 = 32\pi G$ to find that:
\begin{equation}
\frac{d\sigma}{d\Omega} = \frac{16 G^2 m^2}{\theta ^4}.\label{eq:mm4}
\end{equation}
This result exactly matches Eq. \eqref{eq:cross-sec}. Hence we have finished our evaluation of the cross-section using on-shell methods. From Eq. \eqref{eq:mm4}, one has to follow the very same steps detailed in the last part of section \ref{sec:lbqft} in order to obtain the bending angle.

\subsection{Gravitational wave scattering \label{sec:gravscal}}

\subsubsection{Evaluation of the scattering amplitude}

Instead of considering a light ray passing close to a massive object, we could instead imagine a gravitational wave. This corresponds to replacing the external photons with external gravitons. Using traditional quantum field theory methods (i.e. Feynman rules), the 3-graviton vertex is a nightmare inducing 6 index tensor dependant on the three momenta; recall Eq. \eqref{gvertex}. Thankfully for us, we can happily just use little group scaling to derive the 3-point primitives trivially using Eqs. \eqref{3pt1} and \eqref{3pt2}. 

The scattering amplitude that we want to obtain can be represented diagramatically by
\begin{equation}\label{fdiags2}
\begin{gathered}
\begin{fmfgraph*}(120,80)
\fmfleft{i1,i2}
\fmfright{o1,o2}
\fmf{dbl_wiggly,label=$1^{\mp2}$,label.side=left}{i1,v1}
\fmf{plain,label=$4^0$,label.side=left}{i2,v1}
\fmf{dbl_wiggly,label=$2^{\pm2}$,label.side=left}{v1,o1}
\fmf{plain,label=$3^0$,label.side=left}{v1,o2}
\fmfblob{.30w}{v1}
\end{fmfgraph*}
\end{gathered}
\end{equation}

If expanded in terms of Feynman diagrams, this amplitude contains different processes allowed by the interactions in the Lagrangian. However, we can omit this step in the calculation using the BCFW recursion relations.

There is an important difference between this amplitude and the amplitude involving external photons. For the case of photon scattering, scattering amplitudes with helicity configuration $(h_1,h_2)=(\pm1,\pm1)$ for the two photons were identically zero. However, this is not true with gravitons, namely scattering amplitudes with helicity configuration $(h_1,h_2)=(\pm2,\pm2)$ do not vanish. In physical terms, this is due to the fact that a graviton can be absorbed by the scalar and be ejected at some later time, potentially with a different helicity (this process is not allowed for photons, at least for matter that is electrically neutral). However, the physical process we are interested in is the change of angle of a gravitational wave, the wavelength of which is much smaller than the distance to the source of the gravitational field (represente by the scalar field). In this process, the helicity of the wave cannot change. This is the reason why we focus in the following in helicity configurations $(h_1,h_2)=(\mp2,\pm2)$ for external gravitons, as the corresponding amplitudes contain the information about the physical process of interest. The remaining amplitudes would be important to describe other processes, such as the interaction of gravitatonal waves and matter in the early universe, but discussing this phenomenon is out of the scope of this paper.

Hence let us focus on the helicity configuration $(h_1,h_2) = (-2,+2)$, namely we want to evaluate the scattering amplitude $A_4(1^{-2}2^{+2}3^04^0)$. The other helicity configuration can be obtained just by exchanging the two gravitons. The simplest shift we can imagine making is the one where we consider shifting only the two massless gravitons, with external momenta $p_1$ and $p_2$. In this case, it is possible to show using the leading behavior of Feynman diagrams that the shift $[1,2\rangle$-shift is valid for the chosen helicity configuration. Note that we are shifting massless particles (contrary to what we did in the photon case), so the expression of the shift in terms of spinor-helicity variables is simple, and is given by Eq. \eqref{eq:ijshift} with $i=1$ and $j=2$, namely
\begin{equation}
|\hat{1}] = |1] + z|2],~~~~~\ket{\hat{2}} = \ket{2} - z\ket{1}
\end{equation}
Counting the leading $z$-dependence from this Feynman diagram, we find that each vertex contributes a factor $z$, the propagator contributes $z^{-1}$ and the contribution from the each product of polarization vectors is $z^{-2}$. This is sketched in the following figure:
\bigskip
\begin{equation}\label{fdiags2z}
\begin{gathered}
\begin{fmfgraph*}(120,80)
\fmfleft{i1,i2}
\fmfright{o1,o2}
\fmflabel{$z$}{v1}
\fmflabel{$z$}{v2}
\fmflabel{$z^{-2}$}{i1}
\fmflabel{$z^{-2}$}{o1}
\fmf{dbl_wiggly}{i1,v1}
\fmf{plain}{i2,v1}
\fmf{plain,label=$z^{-1}$}{v1,v2}
\fmf{dbl_wiggly}{v2,o1}
\fmf{plain}{v2,o2}
\end{fmfgraph*}
\end{gathered}
\end{equation}
This means that the total diagram has leading dependence $z^{-3}$, to that the shift is valid as explained in section \ref{sec:valid}.

Note that for this shift there are only two possible arrangements of particles that lead to poles, and therefore will contribute to the scattering amplitude as evaluated using BCFW. These two contributions are represented diagrammatically by
\bigskip
\begin{equation}\label{fdiags2}
\begin{gathered}
\begin{fmfgraph*}(120,80)
\fmfleft{i1,i2}
\fmfright{o1,o2}
\fmf{dbl_wiggly,label=$1^{-2}$,label.side=left}{i1,v1}
\fmf{plain,label=$4^0$,label.side=left}{i2,v1}
\fmf{dbl_wiggly,label=$2^{+2}$,label.side=left}{v1,o1}
\fmf{plain,label=$3^0$,label.side=left}{v1,o2}
\fmfblob{.30w}{v1}
\end{fmfgraph*}
\end{gathered}
~~~=~~~
\begin{gathered}
\begin{fmfgraph*}(120,80)
\fmfleft{i1,i2}
\fmfright{o1,o2}
\fmf{dbl_wiggly,label=$\hat{1}^{-2}$,label.side=left}{i1,v1}
\fmf{fermion,label=$4^0$,label.side=left}{i2,v1}
\fmf{plain,label=$\hat{P}_{14}$}{v1,v2}
\fmf{dbl_wiggly,label=$\hat{2}^{+2}$,label.side=left}{v2,o1}
\fmf{fermion,label=$3^0$,label.side=left}{v2,o2}
\end{fmfgraph*}
\end{gathered}
~~~+~~~
\begin{gathered}
\begin{fmfgraph*}(120,80)
\fmfleft{i1,i2}
\fmfright{o1,o2}
\fmf{dbl_wiggly,label=$\hat{1}^{-2}$,label.side=left}{i1,v1}
\fmf{fermion,label=$3^0$,label.side=left}{i2,v1}
\fmf{plain,label=$\hat{P}_{13}$}{v1,v2}
\fmf{dbl_wiggly,label=$\hat{2}^{+2}$,label.side=left}{v2,o1}
\fmf{fermion,label=$4^0$,label.side=left}{v2,o2}
\end{fmfgraph*}
\end{gathered}
\end{equation}

Let us calculate the contributions corresponding to these diagrams using Eq. \eqref{eq:recrel}:
\begin{align}
A_4(1^{-2}2^{+2}3^04^0) &= A_3(\hat{1}^{-2}4^0\hat{P}_{14}^0)\frac{i}{P_{14}^2 +m^2}A_3(\hat{2}^{+2}3^0(-\hat{P}_{14})^0)\nonumber\\
&+A_3(\hat{1}^{-2}3^0\hat{P}_{13}^0)\frac{i}{P_{13}^2 +m^2}A_3(\hat{2}^{+2}4^0(-\hat{P}_{13})^0).\label{eq:zz3}
\end{align}
First of all, let us notice that the only 3-point amplitudes that we need involve two scalar particles and a graviton. These were evaluated in Eqs. \eqref{eq:zz1} and \eqref{eq:zz2}. Hence now it is just a matter of using these expressions and simplifying the answer.

Let us consider the product of 3-point amplitudes in the first line of Eq. \eqref{eq:zz3}. This is given by
\begin{equation}
A_3(\hat{1}^{-2}4^0\hat{P}_{14}^0)A_3(\hat{2}^{+2}3^0(-\hat{P}_{14})^0)=-\frac{\kappa^2}{4}\frac{\langle1|p_4|q]^2\langle g|p_3|2]^2}{[q\hat{1}]^2\langle g\hat{2}\rangle^2}.
\end{equation}
In this expressions, both $|q]$ and $|g\rangle$ can be fixed by choosing a particular gauge (physical results will be independent of this choice). The choice that leads to the simplest result is 
\begin{equation}
|q]=|2],\qquad |g\rangle=|1\rangle,
\end{equation}
which permits to write
\begin{equation}
A_3(\hat{1}^{-2}4^0\hat{P}_{14}^0)A_3(\hat{2}^{+2}3^0(-\hat{P}_{14})^0)=-\frac{\kappa^2}{4}\frac{\langle1|p_4|2]^2\langle 1|p_3|2]^2}{\langle 12\rangle^2[12]^2}=-\frac{\kappa^2}{4}\frac{\langle1|p_4|2]^4}{\langle 12\rangle^2[12]^2}.
\end{equation}
Following the same procedure with the second term in Eq. \eqref{eq:zz3}, we have
\begin{align}
A_4(1^{-2}2^{+2}3^04^0) &=-\frac{i\kappa^2}{4}\frac{\bra{1}p_4|2]^4}{\braket{12}^2[12]^2}\left(\frac{1}{P_{13}^2 +m^2} + \frac{1}{P_{14}^2 +m^2}\right)\\
&=-\frac{i\kappa^2}{4}\frac{\bra{1}p_4|2]^4}{\braket{12}^2[12]^2}\left(\frac{1}{2p_1\cdot p_3} + \frac{1}{2p_1\cdot p_4}\right)\\
&=-\frac{i\kappa^2}{4}\frac{\bra{1}p_4|2]^4}{\braket{12}^2[12]^2}\left(\frac{-2p_1\cdot p_2}{4(p_1\cdot p_3)(p_1\cdot p_4)}\right)\\
&=\frac{i\kappa^2}{16}\frac{\bra{1}p_4|2]^4}{\braket{12}[12]}\left(\frac{1}{(p_2\cdot p_3)(p_2\cdot p_4)}\right)\label{eq:zz4}.
\end{align}

\subsubsection{Evaluation of the cross-section}

As we have done in all the previous examples, we have to obtain the modulus of Eq. \eqref{eq:zz4} in order to extract its physical implications. Proceeding as we did in the photon case and converting to Mandelstam variables, we find that:
\begin{equation}
|A_4(1^{-2}2^{+2}3^04^0)|^2 = |A_4(1^{-1}2^{+1}3^04^0)|^2f(s_{12},s_{13},s_{14})^2,\label{eq:gpprop}
\end{equation}
where we have defined the following function of the Mandelstam invariants:
\begin{equation}
f(s_{12},s_{13},s_{14})=1 - \frac{m^2s_{12}}{(s_{13}+m^2)(s_{14}+m^2)}.\label{eq:propfact}
\end{equation}
Eq. \eqref{eq:gpprop} implies that the cross-section for external gravitons is proportional to the cross-section for external photons, with proportionality factor given in Eq. \eqref{eq:propfact}. From the discussion in section \ref{sec:glgr}, we expect that these two cross-sections should be the same in the geometric optics limit (in which both massless particles follow null geodesics).

Indeed, considering the same approximations as we in the photon case (small deflection angles and large gravitational mass), we find that the function $f(s,t,u)^2 \simeq 1$:
\begin{equation}
f(s_{12},s_{13},s_{14}) \simeq 1-\frac{2 m E \sin ^2\left(\theta/2\right)}{2 m E+4 E^2 \sin ^2\left(\theta/2\right)} \simeq 1.
\end{equation}
Again, in this approximation $E \ll m$ and $\sin^2(\theta/2) \ll 1$. Therefore, we can conclude that the cross-section, and hence the scattering angle, for gravitational and electromagnetic waves is the same in the geometric optics limit.

\section{Multi-Graviton Exchange}\label{sec:Multi}
A natural extension to this work is to consider the contributions to the scattering amplitude by the multi-graviton exchange between the massive object and the scattered photons. Let us recall that in section \ref{sec:lbqft} we argued against the need for these contributions, due to the short time the photon spends in the effective scattering region. This argument holds when attempting to compute the leading order of the amplitude, but due to the long distance nature of gravity there will subleading effects on the photon outside the effective scattering region, which will naturally be suppressed.

These contributions can be computed by means of the so-called ladder diagrams. These diagrams, as will be shown heuristically, form an infinite series which can be resummed in the eikonal approximation. This has been extensively studied in the case of light bending as in \cite{Bjerrum-Bohr:2016hpa}. It should be noted that in these cases standard Feynman techniques are used to determine the resummed form of the contributions offered by the ladder diagrams in the eikonal approximation. A detailed derivation of the resummed ladder diagrams for scalars can be found in \cite{PhysRev.186.1656}, and in \cite{Hinterbichler:2017qyt} the derivation in the appendix is extended to include general particle contents. We will use the case of photon scattering from a massive object to illustrate the idea and then show how modern methods can be used to compute the contributions from these diagrams.

\subsection{Ladder Contributions to Light bending in the Eikonal Limit} \label{sec:lctb}

To compute the formula for the resummed ladder contributions we first need to set up the problem. It makes sense to first consider the eikonal approximation since this will greatly simplify the expressions for the diagrams. The first aspect of the eikonal approximation to consider is kinematic, specifically that the momentum transfer is small $ s_{12} \ll s_{14} $, with momentum labels as used in previous sections. Another way to visualise this is as high energy small angle scattering, that the momentum of the photon in the direction of propagation far exceeds that of the spatially transverse momentum. The next component of the approximation is that we need only consider diagrams of distinct single-graviton exchanges but in all possible permutations, meaning we can disregard any diagrams in which we have multiple gravitons emitted by the scalar at the same point in time.

Now let us build an arbitrary term in the series of ladder diagrams. We have the scalar and photon respectively forming the rails of our ladder and the gravitons the rungs. Consider the $ n- $rung diagram with the $ i' $th graviton having four-momentum $ k_i^\mu $ and the total momentum transfer $ \sum_{i=1}^{n} k_i^\mu = q^\mu $.

\begin{align}
\begin{gathered}
\begin{fmfgraph*}(160,60)
\fmfbottom{i1,d1,d5}
\fmftop{i2,d2,d6}
\fmf{photon}{i1,v1,v2,d5}
\fmf{plain}{i2,v6,v7,d6}
\fmf{dbl_wiggly,tension=0.5,label=$k_1$,label.side=right}{v1,v6}
\fmf{dbl_wiggly,tension=0.5,label=$k_2$,label.side=right}{v2,v7}
\fmflabel{$p_1$}{i1}
\fmflabel{$p_4$}{i2}
\fmfv{label=$p_1+k_1$,label.angle=-60}{v1}
\fmfv{label=$p_4-k_1$,label.angle=60}{v6}
\end{fmfgraph*}
\end{gathered}
\cdots
\begin{gathered}
\begin{fmfgraph*}(40,60)
\fmfbottom{d7,o1}
\fmftop{d8,o2}
\fmf{photon}{d7,v5,o1}
\fmf{plain}{d8,v10,o2}
\fmf{dbl_wiggly,tension=0.5,label=$k_n$,label.side=right}{v5,v10}
\fmflabel{$p_2$}{o1}
\fmflabel{$p_3$}{o2}
\end{fmfgraph*}\nonumber
\end{gathered}
\end{align}

It is necessary to note that we take this diagram to represent all diagrams with $n$ gravitons containing all permutations of the gravitons. It is sufficient to consider only the set of permutations on one of the rails, for example all permutations of the order in which gravitons are absorbed by the photon. Another important point is that this is not a quantum loop diagram but a classical one, due to the limits imposed by the eikonal approximation and the physical considerations: we work in the weak field limit with the photon far from the massive scalar. The amplitude representing this diagram can schematically be expressed as
\begin{equation}
A_n(p_1,p_2,p_3,p_4) \sim \prod_{i=1}^{n} \int d^4 k_i\,\frac{1}{k_i^2 +i\varepsilon} V_{\phi}^i P_{\phi}^i \sum_{\sigma_k}^{ }  V_{\gamma}^i P_{\gamma}^i\,\delta^{(4)}(K - q ),
\end{equation}
where $ P_{\phi}^i  $ and $ V_{\phi}^i $ refer to the scalar propagators and vertices, $ P_{\gamma}^i $ and $ V_{\gamma}^i $ refer to the photon propagators and vertices, $K=\sum_{j=1}^{n}k_j$ and the sum inside the integral is over the set of all permutations $\sigma_k $ of the rungs $ \lbrace k_1, k_2, ... , k_n \rbrace $. Given how we have defined the total momentum transfer, we can take $ p_2 = p_1 + q $. Taking the photon to be initially propagating in the z-direction and squaring this relation we get $2E (-q^0 + q^3) + {\bf q}^2 = 0$, where bold indicates the spatially transverse components. From this we can see that the longitudinal and time components of $ q $ are suppressed by a factor of $ 2E $ in relation to the spatially transverse components, and can be safely disregarded when evaluating the amplitude. This leaves us to deal with the transverse parts of the integrals over $ k_i $ and, due to the symmetries of the vertex expressions, we can drop the sum to pick up a $ 1/n! $ multiplicative factor. As they are unimportant for this discussion, we will suppress any kinematic factors that arise due to the vertices and propagators, since they can always be removed from the integrals in a consistent manner. This leads to an amplitude expression of the form
\begin{align}
A_n(p_1,p_2,p_3,p_4) &\sim \frac{1}{n!} \prod_{i=1}^{n} \int d^2 k_i\,\frac{1}{{\bf k}_i^2 + i\varepsilon} \delta^{(2)}( {\bf K} - {\bf q}) \nonumber \\
&=  \int d^2 b\, e^{-i {\bf b}\cdot {\bf q} } \frac{1}{n!} \left[ \int d^2 k\, \frac{1}{{\bf k}^2 +i\varepsilon} e^{ i {\bf b}\cdot{\bf k} } \right]^n \nonumber\\
&= \int d^2 b\, e^{-i {\bf b}\cdot {\bf q} } \frac{1}{n!} \chi(\textbf{b})^n .
\end{align}
In the equation above, we have expressed the delta function in integral form as
\begin{equation} 
\delta^{(2)} \left( {\bf K} - {\bf q} \right) =  \int d^2 b\, e^{-i {\bf b}\cdot \left( \sum_{j=1}^{n}{\bf k}_j - {\bf q} \right) },
\end{equation}
and labelled the factor in square brackets as $ \chi({\bf b}) $. 

We can now perform a Fourier transform for the $ n=1 $ case (the tree-level amplitude) between $ q $ and its dual variable $ b $, which is nothing but the impact parameter. This immediately leads to an expression for $ \chi({\bf b}) $ in terms of the tree level amplitude,
\begin{equation}
\chi({\bf b}) \sim \int d^2 q\, e^{i {\bf b}\cdot {\bf q} } A_1(p_1,p_2,p_3,p_4).
\end{equation}

We can now sum over all orders in the ladder diagrams, which amounts to summing the amplitude $ A_n(p_1,p_2,p_3,p_4) $ over all $n$. It should be apparent that this is simply the Taylor expansion of an exponential
\begin{equation}
A^{\rm ladder}(p_1,p_2,p_3,p_4) = \sum_{n} A_n(p_1,p_2,p_3,p_4) \sim \int d^2b\, e^{-i {\bf b}\cdot {\bf q} } \left[e^{\chi({\bf b})} -1\right].
\end{equation}
This gives us a remarkably simple formula for the fully resummed ladder diagrams, written only in terms of the tree level amplitude. This allows us to compute additional contributions to a given event with relative ease.

This method to derive this formula is well established in QFT, however very little effort (as far as we know) has been made to recast this in a modern amplitude framework or to apply it to interactions involving gravitational waves. In the next subsection, we will discuss some of the directions that could be taken.

\subsection{A Modern Approach to Ladders}\label{sec:ModernLadders}
Modern amplitude methods are particularly well suited to problems involving multiple interactions. Naturally we want to utilise these methods to tackle the scenarios presented earlier in the paper. As a starting point, we can consider the ladder diagrams in the eikonal approximation, as we have already seen that this greatly limits the number of diagrams we need to consider. To this end, we first need to introduce the idea of generalised unitarity as a method to compute loop diagrams (even if these are classical loops). For a more detailed introduction of the method see \cite{Elvang:2013cua} and \cite{Henn:2014yza}.

It is well established that if one has a generic one-loop amplitude, in integral form, it can be expressed as a linear combination of scalar \textit{master integrals} involving the box, triangle, bubble and tadpole diagrams. This is due to the fact that (at one loop) products of external momenta and internal loop momenta can always be written in terms of propagators, a technique that is often referred to as Passarino-Veltman reduction \cite{Passarino:1978jh}. For the purpose of illustrating the techniques here we will restrict ourselves to a generic four-point one-loop amplitude $ A_4^1 $ in four dimensions. This reduction is expressed as 
\begin{equation}
A_4^1 = \sum_i \left[ c_4^i I_4^i + c_3^i I_3^i + c_2^i I_2^i +c_1^i I_1^i \right] +\mathcal{R},
\end{equation}
where the sum is over all the possible configurations of the external momenta. The $ c_j $'s are coefficients consisting of kinematic invariants and the $ I_j $'s are the scalar integrals, and $\cl{R}$ is a rational part that typically a remnant of dimensional reduction, which we assume throughout this section. For example, $I_4$ is the scalar box integral given by
\begin{align}
&I_4(\lbrace p_i^2 \rbrace; s_{12},s_{23}; \lbrace m_i^2 \rbrace) =
\begin{gathered}
\begin{fmfgraph*}(100,60)
\fmfleft{i1,i2}
\fmfright{o1,o2}
\fmf{plain}{i1,v1,v2,o1}
\fmf{plain}{i2,v3,v4,o2}
\fmf{plain}{v1,v3}
\fmf{plain}{v2,v4}
\fmflabel{$p_4$}{i2}
\fmflabel{$p_1$}{i1}
\fmflabel{$p_2$}{o1}
\fmflabel{$p_3$}{o2}
\fmfv{label=$l_2$,label.angle=-120}{v2}
\fmfv{label=$l_1$,label.angle=120}{v1}
\fmfv{label=$l_3$,label.angle=-60}{v4}
\fmfv{label=$l_4$,label.angle=60}{v3}
\end{fmfgraph*}\nonumber
\end{gathered} \nonumber \\\nonumber\\
&=
\int \frac{d^{4-2\varepsilon} l_1}{(2\pi)^{4-2\varepsilon}} \frac{1}{(l_1^2+m_1^2-i\varepsilon)(l_2^2+m_2^2-i\varepsilon)(l_3^2+m_3^2-i\varepsilon)(l_4^2+m_4^2-i\varepsilon)}.
\end{align}
Here, the $ m_i $'s are the masses of the internal lines. 

This technique of integral reduction reduces the problem of computing one-loop amplitudes with complicated tensor structures\footnote{By complicated tensor structure' we mean a loop integrand with some numerator involving products of internal and external momenta, polarization vectors/tensors etc. Scalar loop integrands conveniently do not have such annoying features.} to determining the coefficients of the various scalar integrals and the rational part. To compute the coefficients, we turn to the very efficient method of generalised unitarity, which in essence is a generalisation of the optical theorem, which relates different orders in perturbation theory. We can then efficiently compute loop amplitudes by studying the discontinuities of the various kinematic channels, and utilising what we know about the factorisation property of amplitudes.

The optical theorem in QFT uses the unitary nature of the S-matrix, $ SS^\dagger = 1 $. We expand the S-matrix into its trivial and non trivial part, $ S=1+iT $. Taken together, and written in terms of matrix elements, we find
\begin{equation}\label{eq:opth}
i ( T - T^\dagger) = T T^\dagger \implies i\braket{f|T|i} - i\braket{f|T^\dagger|i} = \int d\mu \braket{f|T|\mu}\braket{\mu|T^\dagger|i}.
\end{equation}
This is known as the generalised optical theorem, and we can see exactly how it relates the contributions of $T$ at different orders once one considers the perturbative expansion of $T$ in terms of the coupling constant $g$,
\begin{equation}
T = g^2 T^{\rm(Tree)} + g^4 T^{\rm (1-loop)}  + O(g^6).
\end{equation}
Plugging this in equation \eqref{eq:opth}, we immediately see that
\begin{equation}
i ( T^{\rm(1-loop)} - T^{\rm(1-loop)\dagger}) = T^{\rm(Tree)} T^{\rm(Tree)\dagger}.
\end{equation}

This is known as Cutkosky's rule \cite{Cutkosky:1960sp}, which allows us to express loop amplitudes in terms of tree-level amplitudes. We can represent this diagrammatically as
\begin{align}
&\begin{gathered}
\begin{fmfgraph*}(60,40)
\fmfbottom{i1,d1,o1}
\fmftop{i2,d2,o2}
\fmf{dashes}{d1,d2}
\fmf{plain}{i1,v1,v2,o1}
\fmf{plain}{i2,v3,v4,o2}
\fmf{plain}{v1,v3}
\fmf{plain}{v2,v4}
\fmflabel{$p_4$}{i2}
\fmflabel{$p_1$}{i1}
\fmflabel{$p_2$}{o1}
\fmflabel{$p_3$}{o2}
\fmfv{label=$l_2$,label.angle=-120}{v2}
\fmfv{label=$l_1$,label.angle=120}{v1}
\fmfv{label=$l_3$,label.angle=-60}{v4}
\fmfv{label=$l_4$,label.angle=60}{v3}
\end{fmfgraph*}
\end{gathered} 
\nonumber\\
&= \sum_{\rm Internal\ states} \int d^4 l_2 \delta (l_2^2) \delta [(l_2-p_1-p_2)^2]
\begin{gathered}
\begin{fmfgraph*}(60,40)
\fmfleft{i1,i2}
\fmfright{o1,o2}
\fmf{plain}{i1,v1,o1}
\fmf{plain}{i2,v3,o2}
\fmf{plain}{v1,v3}
\fmflabel{$p_1$}{i1}
\fmflabel{$p_4$}{i2}
\fmflabel{$-l_4$}{o2}
\fmflabel{$l_2$}{o1}
\end{fmfgraph*}
\end{gathered} 
~~~\times~~~
\begin{gathered}
\begin{fmfgraph*}(60,40)
\fmfleft{i1,i2}
\fmfright{o1,o2}
\fmf{plain}{i1,v1,o1}
\fmf{plain}{i2,v3,o2}
\fmf{plain}{v1,v3}
\fmflabel{$-l_2$}{i1}
\fmflabel{$l_4$}{i2}
\fmflabel{$p_3$}{o2}
\fmflabel{$p_2$}{o1}
\end{fmfgraph*}
\end{gathered} 
\end{align}
In essence what the cut does is force the cut internal momenta  $ l_2 $ and $ l_4 $ to go on-shell, giving rise to a discontinuity around which the amplitude can be factorised. This diagram only represnets the s-channel cut of the box diagram: to get the complete box contribution one must also perform a u-channel cut, in other words we also have to cut the horizontal internal momenta. To get the complete loop amplitude one must then also consider the possible triangle, bubble and tadpole diagrams.

Luckily, we are interested in determining the first loop term in the series of ladder diagrams in the eikonal limit, meaning we need only consider the box diagram contribution. For the case of light bending due to a massive scalar, we cut the following diagram along the s-channel forcing $ l_2 $ and $ l_4 $ to go on-shell:

\begin{align}
\begin{gathered}
\begin{fmfgraph*}(160,80)
\fmfbottom{i1,d1,o1}
\fmftop{i2,d2,o2}
\fmf{photon}{i1,v1}
\fmf{photon}{v1,v2}
\fmf{photon}{v2,o1}
\fmf{plain}{i2,v3}
\fmf{plain}{v3,v4}
\fmf{plain}{v4,o2}
\fmf{dbl_wiggly,tension=0.5}{v1,v3}
\fmf{dbl_wiggly,tension=0.5}{v2,v4}
\fmf{dots}{d1,d2}
\fmflabel{$4^0$}{i2}
\fmflabel{$1^{+1}$}{i1}
\fmflabel{$2^{-1}$}{o1}
\fmflabel{$3^0$}{o2}
\fmfv{label=$l_2$,label.angle=-120}{v2}
\fmfv{label=$l_1$,label.angle=120}{v1}
\fmfv{label=$l_3$,label.angle=-60}{v4}
\fmfv{label=$l_4$,label.angle=60}{v3}
\end{fmfgraph*}\nonumber
\end{gathered}
\end{align}

In terms of the on-shell amplitudes, this can be written as
\begin{align}
&\int d^4 l_2 \delta (l_2^2) \delta [(l_2-p_1-p_2)^2] \left[ A_4^{\rm Tree}(1^{+1},l_2^{-1},-l_4^0,4^0) A_4^{\rm Tree}(-l_2^{+1},2^{-1},3^0,l_4^0) \right]\nonumber\\
&= c_4^s\int d^4 l_2 \delta (l_2^2) \delta [(l_2-p_1-p_2)^2] \frac{1}{l_3^2l_1^2}
\end{align}
Since the integral measure in both the left and right hard sides are the same we can focus on the integrand itself. Notice that there is only one possible way we can organise the internal states for the given cut, since any vertex with two photons of the same helicity is zero. Recycling the expressions we derived for the four-point tree amplitudes of this form, equation \eqref{eq:nn4}, we can write the above expression as
\begin{equation}
\frac{\kappa^4}{16} \frac{\bra{l_2}p_1|1]^2}{(l_2+p_1)^2} \frac{\bra{2}p_4|l_2]^2}{(l_2-p_2)^2} = c_4^s \frac{1}{(l_2+p_1)^2}\frac{1}{(l_2-p_2)^2}.
\end{equation}
Comparing the left and right sides of the expression we find a simple relation for the $ c_4^s $, we just need to remove the $ l_2 $ dependence from the expression. To this end recall that $ p_1+l_2-l_4+p_4 =0 $ and $ p_2-l_2+l_4+p_3 =0 $ by conservation of momentum. Using this we find
\begin{align}
c_4^s &= \frac{\kappa^4}{16} (\bra{2}p_3|l_2] \bra{l_2}p_4|1]) ^2 \nonumber\\
&= \frac{\kappa^4}{16} (\bra{2}l_4|l_2] \bra{l_2}p_4|1]) ^2  \nonumber\\
&= \frac{\kappa^4}{16} (\bra{2} l_4 l_2 p_4|1]) ^2 \nonumber\\
&= \frac{\kappa^4}{16} (\bra{2} l_4 (l_4 - p_1 - p_4) p_4|1]) ^2.
\end{align}

Now we can impose the kinematic limits of the eikonal approximation, namely that the momentum transfer is small, along with the assumption that the mass of the scalar is much larger than the other energy scales in the problem. This allows us to write that $ l_4 \simeq p_3 \simeq p_4 $. Firstly this lets us write the $ l_4 - p_1 - p_4 \simeq - p_1 $ and also allowing $ \bra{2}l_4|1] \simeq \bra{2}p_4|1]$. We also know that $ s_{14} = -m^2 + \bra{1}p_4|1] $. Finally then, the coefficient is given by

\begin{equation}
c_4^s = \frac{\kappa^4}{16} \bra{2}p_4|1]^2 (s_{14}+m^2)^2.
\end{equation}

This brings the first loop term of the ladder series to the form

\begin{equation}
A_4^{\rm 1-loop} (1^{+1}2^{-1}3^0 4^0) = \frac{\kappa^4}{16} \bra{2}p_4|1]^2 (s_{14}+m^2)^2 I_4(s_{14},s_{23}),
\end{equation}

plus some other terms with alternate external momentum configuration, $ p_3 \leftrightarrow p_4 $, which can be handled in a similar way. This can be simply extended to the case involving external gravitons, and it is hopeful that by recursively adding tree diagrams to this loop diagram by means of generalized unitarity that we can build the full series of ladder contributions.

 \section{Conclusions}\label{sec:con}
Adopting the philosophy that many astrophysical processes can be viewed as scattering problems, this article is our modest attempt at a pedagogical account, to an astrophysics audience, of the so-called\footnote{By us, admittedly.} amplitude revolution sweeping across the landscape of modern quantum field theory. In it we have introduced a number of concepts. These include the spinor-helicity formalism of square and angled bra-kets, the KLT factorization of gravity into the square of a gauge theory and, most importantly, the BCFW recursion relations and exemplified their use through one of the most iconic problems in gravitational physics, namely gravitational lensing. We certainly anticipate that much of this mathematical machinery will be new to our intended readership. But then, it is not our intension to have 
the reader emerge at this point proficient in computing astrophysical amplitudes with the BCFW relations or even in using spinor-helicity variables in their everyday lives\footnote{Although we would be delighted if they do.}. Instead, our aim in this first article is merely to convey the potential power of the formalism. 

Now, like using a sledgehammer to crack a nut, it might seem that applying the amplitude formalism to light-bending is, well, overkill. This is true. However, it also true that we find ourselves on the precipice of a new era in astrophysics, an era of big data\footnote{No, we don't really know what this is either.}, of precision observation and, perhaps most excitingly, an era of gravitational wave astronomy. And this is where we expect amplitude methods to come into their own. As a set of tools developed in the laboratories of high energy physics, computing quantum corrections to classical results comes naturally to it. So while the corrections to light-bending etc. may be small (see, for example, \cite{Bjerrum-Bohr:2014zsa} for an excellent treatment of this issue), it is eminently calculable in the scattering formalism. Moreover, as we hope to have convinced the reader in section 4.2 and will say more about in a forthcoming work, scattering gravitons off massive bodies (as one would expect to do in studying gravitational wave physics) is not much more difficult than scattering photons. Actually, while we have restricted our attention to {\it astrophysics} in this article, these methods have also been adapted quite successfully to more {\it cosmological} questions. For example,   
the introduction of spinor-helicity variables was key to describe circularly polarized gravitational waves in de Sitter space in \cite{Maldacena:2011nz}. This in turn was required in order to compute the graviton 3-point function that contributed to graviton non-Gaussianities during inflation. Again, while the effect is small, the very idea of such a detailed analysis of the tensor fluctuations during inflation was unheard of only a few years ago. 

In as much as we focussed on demonstrating the efficacy of a new set of mathematical tools from the high energy toolkit to an astrophysical audience, we also hope to also have pointed out to a more high energy readership, a different set of problems, both new and old, worth contemplating. In any event, we hope that this is not so much a conclusion as it is an introduction.

\section*{Acknowledgements}

We would like to thank Timothy Adamo, Daniel Grin, Malcolm Perry and Edward Witten for useful discussions. RCR is funded by a postdoctoral fellowship from the Claude Leon Foundation of South Africa. DB is supported by a PHD fellowship from the South African National Institute for Theoretical Physics (NITheP). JM gratefully acknowledges support by NSF grant PHY-1606531 at the Institute for Advanced Study and NRF grant GUN 87667 at the University of Cape Town. AW would like to thank the Institute for Advanced study, the Simons Foundation and the Flatiron Institute where part of this work was completed, as well as the Princeton University Astrophysics department for their generous support. AW and NM are supported by the South African Research Chairs Initiative of the Department of Science and Technology and the National Research Foundation of South Africa. Any opinion, finding and conclusion or recommendation expressed in this material is that of the authors and the NRF does not accept any liability in this regard. Finally, JM and AW would like to thank Carl Feinberg for his generous support of the accumulation of useless knowledge at the Institute for Advanced Study.

\appendix
\section{Some group theory}\label{sec:group_theory}

A {\bf group}, as everyone knows, is a set of objects closed under a group operation that satisfies the axioms of {\it identity, associativity} and {\it invertibility}. This definition is an abstract one, unmarried to any particular realization of the objects in the set. 
For our purposes, it will suffice to take a representation of $G$ as a set of {\it matrices}. In this case, group multiplication is synonymous with standard matrix multiplication. In this article, we are mostly interested in {\bf Lie groups}, G say, which are groups associated to some continuous set of parameters that in turn can be taken as coordinates in some continuous manifold, the Lie group manifold. 

For an example, one need look no fourther than the group of {\it rotations} on the plane where a 2-vector $\vec{v}$, is transformed into the vector $\vec{v'}$ by the action of a $2\times 2$ matrix,
\begin{equation}
\vec{v}' = R\vec{v},
\end{equation}
which depends on the (continuous, periodic) rotation angle $\theta$. This is the Lie group $O(2)$, the set of $2\times 2$ orthogonal matrices. In general this group contains not only rotations but also reflections. We can restrict this group to one that does not contain reflections by demanding that $\det R = +1$. This restricted group is the {\bf special orthogonal group} $SO(2)$. It is not hard to see that this concept is easily generalised to rotations in $N$ dimensions where,
\begin{equation}
SO(N) = \left\{R_{N\times N}~\bigg|~ R^TR = 1 ~~\text{and}~~\det R = 1\right\}\,.
\end{equation}

As is often the case in physics, it is often more convenient to work with {\it infinitesimal} transformations {\it i.e.} to ``Taylor expand" transformations about the identity. To this end let's look at an infinitesimal rotation parameterized by an angle $\alpha$. In this case, the rotation matrix is ``close" to the identity, so we can write:
\begin{equation}\label{key}
R \simeq I + A\,.
\end{equation}
Since $R$ is a rotation matrix, $R^TR = (I + A^T)(I + A) = I + A^T + A = I$. For this to hold, $A$ must be an antisymmetric matrix, {\it i.e.} $A^T = -A$. In 2 dimensions, this means that 
\begin{equation}
A = \alpha J = \alpha\begin{pmatrix}[cc]
0 & 1 \\ 
-1 & 0
\end{pmatrix}\,.
\end{equation}
The $2\times2$ matrix $J$ is known as the {\bf generator} of the group. 

In order to relate this back to a finite angle $\theta$, we replace $\alpha \rightarrow \theta/N$ and apply the rotation $N$ times, finally taking the $N\rightarrow\infty$ limit\footnote{To check explicitly that this gives the rotation matrix, do a taylor expansion of the exponential using the standard formula},
\begin{equation}
R(\theta) = \lim_{N\rightarrow\infty}\left(I + \frac{\theta J}{N}\right)^N = \exp(\theta J)\,.
\end{equation}
In the case of the $N$ dimensional rotation group $SO(N)$, this becomes $R = \exp(\theta_a J^a)$.

The number of real parameters $n$ that characterise this group defines the {\bf dimension} of the group, and is equal to the number of independent elements of the group generators. For $SO(N)$, the number of independent elements in an antisymmetric $N\times N$ matrix
\begin{equation}\label{dimG}
n = \dim{SO(N)} = \frac{N(N-1)}{2}\,.
\end{equation}
A closely related, and frequently occuring, group is the {\bf special unitary} group $SU(N)$, which is the set of $N\times N$ unitary matrices with unit determinant. Following the same arguments as for the special orthogonal group, we can show that this group has $N^2 -1$ generators. This group can be thought of as rotations in $N$ \textit{complex} dimensions.

The utility of infinitesimal transformations has already manifest in our discussion. Geometrically, restricting to transformations close to the identity corresponds to working in the {\it tangent vector} space at the origin of the group manifold defines. For a given Lie group $G$, this defines the associated\footnote{There are many differing notations for the Lie algebra associated to a particular group. Another popular choice is $\mathfrak{g}$.} {\bf Lie algebra}, $g$ with the group generators forming a basis for the vector space. Equivalently, the algebra is defined by the commutation relation satisfied by the group generators,
\begin{equation}
[T^a,T^b] = f^{ab}_cT^c\,.
\end{equation}
Here, the $f^{ab}_c$ are the {\bf structure constants} of the Lie algebra. Given a Lie algebra, an element of the associated group is found by exponentiating an element of the algebra, as for the rotation group. Schematically,
\begin{center}
	Lie Algebra $~~\xrightarrow{exp}~~$ Lie Group\,.
\end{center}
For the groups $SO(3)$ and $SU(2)$, note that while the generators are different, the structure constants of each of these groups is the same, $f_{abc} = \epsilon_{abc}$, the completely antisymmetric tensor in 3 dimensions. Consequently, the structure of each group must be connected in some way. In fact, both $SO(3)$ and $SU(2)$ are rotation groups with 3 real parameters.

One dimension down, the group $SO(2)$ acts on 2-vectors $\vec{v}$. However we can obviously rotate higher dimensional vectors in the same 2-plane as well\footnote{In 3 dimensions, imagine transforming only the $x$ and $y$ components of a vector in $\R^3$, leaving the $z$ component untouched.}. This is where the idea of a {\bf group representation} enters. Matrices that act on a particular space that have the same structure as the group, but where the matrices can have a rank different from $N$. In practice, this means that the group generators will be matrices of rank $M$ that obey the group algebra (and have the same structure constants). We can classify some interesting representations using $N$ and $M$:

\begin{itemize}
	\item The {\bf trivial} representation is one where $M = 1$. The generators are scalars, and nothing 
	happens to vectors under group transformations.
	\item The {\bf fundamental} representation is where $M = N$. In this case, the generators 
	are themselves the group matrices, i.e. the $SO(N)$ rotations acting on $\R^N$. Elements of the
	group are vectors.
	\item The {\bf adjoint} representation is where $M = \dim{G}$. In this representation the 
	structure constants generate the group and group elements are represented as matrices.
\end{itemize}
Taking $SO(3)$ as an example, eq. \eqref{dimG} tells us that the group acts naturally on a 3 dimensional vector space spanned by the 3 generators $T^a$ of the adjoint representation.

\subsection{$SO(3)$ and $SU(2)$}

Let's expand a little on some of the groups that feature prominently in this article. We have already seen that $SO(3)$ is the group of orthogonal $3\times 3$ matrices with unit determinant. The group $SU(2)$ is defined similarly, as $2\times 2$ complex, unitary matrices with unit determinant,
\begin{equation}
SU(2) = \left\{\begin{pmatrix}
\alpha & \beta^*\\
\beta & -\alpha
\end{pmatrix}~~\Bigg|~~ |\alpha|^2 + |\beta|^2 = 1\right\}\,.
\end{equation}
Here $\alpha$ is a real number and $\beta$ is complex. Consequently, each $SU(2)$ transformation depends on 3 real parameters. Elements of $SU(2)$ can be neatly expanded in a basis of {\bf Pauli matrices},
\begin{equation}\label{paulim}
\sigma_1 = \begin{pmatrix}
0 & 1\\
1 & 0
\end{pmatrix}\,,
\qquad
\sigma_2 = \begin{pmatrix}
0 & -i\\
i & 0
\end{pmatrix}\,,
\qquad
\sigma_3 = \begin{pmatrix}
1 & 0\\
0 & -1
\end{pmatrix}\,.
\end{equation}
This means that any $SU(2)$ matrix can be written as $M = \theta^a\sigma_a/2$, and therefore a group element can be written $U=\exp(i\theta^a\sigma_a/2)$. The three real components of an element of $SU(2)$ uniquely determine a unit 3-vector $\vec{v} = (x,y,z)$, and transformations under $SU(2)$ lead to another unit 3-vector $\vec{v}' = (x',y',z')$. This means that $SU(2)$ constitutes a rotation in 3 dimensions, exactly like $SO(3)$. In fact, $SU(2)$ is the so-called (double) covering group of $SO(3)$.

\subsection{The Lorentz group}
The Lie group of most interest to us is the {\bf Lorentz group} of rotations and boosts in 4-dimensional Minkowski space. For our purposes though, it will suffice to consider the subset of proper orthochronus Lorentz transformations, that exclude the discrete parity and time-reversal transformations. These are elements of a restricted Lorentz group $SO^+(1,3)$, where the plus means 'restricted', and is written here only once and implied for the rest of this section. Specifically, Lorentz transformations act on the spacetime metric $g_{\mu\nu}$ as,
\begin{equation}
\Lambda^T g_{\mu\nu} \Lambda = g_{\mu\nu}\,.
\end{equation}
Close to the identity, the transformations can be expanded as
\begin{equation}
\Lambda = I + \frac12\omega_{\mu\nu}M^{\mu\nu}\,,
\end{equation}
where $M_{\mu\nu}$ is the generator of the group, defined as $M_{\mu\nu} = \partial_\mu x_\nu - \partial_\nu x_\mu$ and $\omega_{\mu\nu}$ parameterize the transformations.
The generators satisfy the Lie algebra:
\begin{equation}
[M_{\mu\nu},M_{\rho\sigma}] = -i(g_{\mu\rho}M_{\nu\sigma} - g_{\mu\sigma}M_{\nu\rho} +  
g_{\nu\sigma}M_{\mu\rho} - g_{\mu\rho}M_{\nu\sigma}).
\end{equation}
These generators as defined above are a convenient notation that captures both the usual rotation generators $J_i$ that we have encountered already, together with the generators $K_i$ of Lorentz boosts through
\begin{equation}
J_i = \frac12 \epsilon_{ijk}M_{jk},~~~~\text{and}~~~~K_i = M_{0i}\,.
\end{equation}
These generators satisfy their own algebras, one of which we know as the $SO(3)$ algebra,
\begin{equation}
[J_i,J_j] = i\epsilon_{ijk}J_k,
\end{equation}
while 
\begin{eqnarray}
[J_i,K_j] = i\epsilon_{ijk}K_k\,,~~~~~~[K_i,K_j] = -i\epsilon_{ijk}J_k\,.
\end{eqnarray}
This furnishes one representation of the Lorentz group. It will prove instructive to look for another by changing basis. To this end, let's define two new operators from linear combinations of boosts and rotations,
\begin{equation}\label{key}
J^\pm_i = \frac12 (J_i \pm iK_i)\,.
\end{equation} 
Each of these are independent, a fact easily demonstrated by checking that $[J_i^+,J_j^-] = 0$. 
Further, by substituting in the relevant operators, it can be shown that
\begin{equation}
[J^\pm_i,J^\pm_j] = \frac12 \epsilon_{ijk}J_k^\pm\,.
\end{equation}
This is precisely the algebra of the rotation group and $SU(2)$, except now there are two copies, one for the `+' and one for the `-' generators. In other words, we have arrived at another representation of the Lorentz group where evidently
\begin{equation}
SO(1,3)\simeq SU(2)\otimes SU(2)\,.
\end{equation}
To finish off this lightning review of group theory, let's discuss one more important group that is a cover of the Lorentz group. This is the complex special linear group $SL(2,\C)$. This group is very similar to $SU(2)$, except that {\it all} entries of the $2\times 2$ matrices are complex,
\begin{equation}
SL(2,\C) = \left\{\begin{pmatrix}
a & b\\
c & d
\end{pmatrix}~~\Bigg|~~ ad - bc = 1\right\}\,.
\end{equation}
where $a,b,c,d$ are now complex numbers. Clearly $SU(2)$ is a subgroup of $SL(2,\C)$. 

Any vector $V^\mu$ in $\R^{1,3}$ can be represented as a matrix in $SL(2,\C)$, since we can encode the information of a Lorentz 4 vector as:
\begin{equation}
V^\mu = (t,x,y,z) \longrightarrow W = \begin{pmatrix}
t + z & x - iy\\
x + iy & t - z
\end{pmatrix} = tI_{2\times 2} + ix^i\sigma_i\,.
\end{equation}
If we define the sigma matrices as,
\begin{equation}\label{key}
\sigma^\mu = (I,\sigma^i),~~~~~~~~~~~~\overline{\sigma}^\mu = (I,-\sigma^i)\,,
\end{equation}
then we can define the standard conversion from a Lorentz four vector to a $2\times 2$ matrix of $SL(2,\C)$ via the new basis\footnote{$\sigma^\mu$ represents a basis of $SL(2,\C)$.}:
\begin{equation}
V^\mu \longrightarrow W = V^\mu\sigma_\mu
\end{equation}  
The square of a four vector in $\R^{1,3}$ is equal to the determinant of $W$.

\section{Scattering Cross-section} \label{sec:scs}

In section \ref{sec:lbqft} we relate the scattering amplitude to the differential cross-section and this in turn to the impact parameter in order to calculate the scattering angle. Here, we give a brief argument to illustrate how one arrives at these equations. To relate the cross-section to the scattering amplitude we consider a scattering event with states $ \phi_i(\bar{p}_i) $. For simplicity, we limit ourselves to a two to two scattering process with labels as used in the rest of this article: $ \phi_1 \phi_4 \rightarrow  \phi_2 \phi_3$. We specify no particle content for the derivation that follows and institute the appropriate limits at the end, closely following the treatment given in \cite{Peskin:1995ev}. All the states are appropriately normalised and we suppose that the initial states are very closely distributed, in momentum space, around some central momenta, say $ p'_1 $ and $ p'_4 $. The initial state of our system is given by

\begin{equation}
\vert \phi_1(p_1) \phi_4(p_4) \rangle = \int \frac{d^3 \bar{p}_1 d^3 \bar{p}_4}{(2 \pi)^6 \sqrt{4 E_1 E_4}} \phi_1(\bar{p}_1) \phi_4(\bar{p}_4) e^{-i \bar{b}\cdot \bar{p}_1} \vert \bar{p}_1 \bar{p}_4 \rangle .
\end{equation}

Of course any experiment requires repetition so we can consider many states of type $ \phi_1 $ in a cylindrical beam along the z-axis incident on a single target state $ \phi_4 $ situated in the centre of the beam. We will assume that the incident particles are uniformly distributed about the z-axis and that the transverse offset is given by the impact parameter $ b $ radially from the centre of the beam. This means that the exponential factor in the above equation will account for this offset. Since the differential cross-section is an infinitesimal quantity, we only want to consider final states that fall within some small measure of momentum space, i.e. $ d^3\bar{p}_3 d^3\bar{p}_2 $. We are interested in the scattering probability 

\begin{equation}
\mathcal{P}(p_1,p_4\rightarrow p_3,p_2) = \frac{d^3 \bar{p}_3 d^3 \bar{p}_2}{(2 \pi)^6 4 E_3 E_2} \left\vert \langle \bar{p}_3 \bar{p}_2 \vert \phi_1(p_1) \phi_4(p_4) \rangle \right\vert^2 .
\end{equation}

The cross-section $ \sigma $ is defined as the ratio 
\begin{equation}\label{key}
\sigma = \frac{N}{n_i\times n_p}
\end{equation}
Where $N$ is the number density of scattering events, $n_i$ is the number of incident particles and $n_p$ is the number of particles at the scattering centre, in our case $n_p = 1$. The number of scattering events can be written as the number of incident particles times the probability that a scattering event will occur within some region of the beam,

\begin{equation}
N = \int d^2b\, n_i \mathcal{P}(\bar{b}) .
\end{equation}

The cross-section is then

\begin{equation}
\sigma = \frac{N}{n_i \times 1} = \int d^2 b\, \mathcal{P}(\bar{b}),
\end{equation}

and using the previous expressions we write the infinitesimal cross-section as

\begin{equation} \label{eq:diffcross}
d\sigma = \frac{d^3 \bar{p}_3 d^3 \bar{p}_2}{(2 \pi)^6 4 E_3 E_2} \int d^2 e^{i \bar{b}\cdot (\bar{p}_1^*-\bar{p}_1)} 
\left\vert \int \frac{d^3 \bar{p}_1 d^3 \bar{p}_4}{(2 \pi)^6 \sqrt{4 E_1 E_4}} \phi_1(\bar{p}_1) \phi_4(\bar{p}_4) \langle \bar{p}_3 \bar{p}_2 \vert \bar{p}_1 \bar{p}_4 \rangle \right\vert^2 .
\end{equation}

To simplify the above expression we first perform the integral over $ b $ to get a delta function $ (2\pi)^2\delta^{(2)}(\tilde{p}^*_1-\tilde{p}_1) $, where the tilde indicates the directions transverse to the z-axis. We also write the correlation function in terms of the scattering amplitude

\begin{equation}
\langle \bar{p}_3 \bar{p}_2 \vert \bar{p}_1 \bar{p}_4 \rangle = (2\pi)^4 \delta^{(4)} (p_1+p_4 - p_3-p_2) A(p_1,p_2,p_3,p_4),
\end{equation}

providing some more delta functions. This allows us to do all the integrals in the complex conjugate part of equation \eqref{eq:diffcross}, after which we can fix the incident states' momenta to the central momenta mentioned at the start of the appendix. Therefore any smooth functions dependent on $ \lbrace p_1, p_4 \rbrace $ can be evaluated at $ \lbrace p'_1, p'_4 \rbrace $ and removed from the integral. If we also assume that the resolution of the detector is insufficient to resolve the fluctuations of the initial state momenta from the central value we can set all the remaining momenta to the central value. This gives the expression

\begin{align}
d\sigma &= \frac{1}{4 E_1 E_4 \vert v_1-v_4 \vert}\frac{d^3 \bar{p}_3 d^3 \bar{p}_2}{(2 \pi)^6 4 E_3 E_2}  
\vert A(p'_1,p_2,p_3,p'_4) \vert^2 (2\pi)^4 \delta^{(4)} (p'_1+p'_3 - p_3-p_2) \nonumber\\
&\times \int \frac{d^3 p_1 d^3 p_3}{(2\pi)^6} \vert \phi_1(\bar{p}_1) \vert^2 \vert \phi_4(\bar{p}_4) \vert^2 \nonumber\\
d\sigma &= \frac{1}{4 E_1 E_4 \vert v_1-v_4 \vert}\frac{d^3 \bar{p}_3 d^3 \bar{p}_2}{(2 \pi)^6 4 E_3 E_2}  
\vert A(p'_1,p_2,p_3,p'_4) \vert^2 (2\pi)^4 \delta^{(4)} (p'_1+p'_4 - p_3-p_2),
\end{align}

where $ v_1,v_4 $ are the velocities of the initial states along the z-axis. To get the above expression in a form we can use we need to partially evaluate the phase-space integrals on the right hand side. We can compute the integral over $ \bar{p}_3 $ using the delta function corresponding to the 3-momentum and we can switch to spherical coordinates in order to do the integral over $ \bar{p}_2 $, where the solid angle is $ d\Omega = \sin(\theta) d\theta d\phi $, to find in the end

\begin{equation}
I = \int \frac{ d\Omega}{16 \pi^2} \frac{\vert\bar{p}_2\vert}{E_1+E_4}.
\end{equation}

Inserting this into the latest expression for the cross-section and rearranging we have

\begin{equation} \label{eq:diffcross2}
\frac{d\sigma}{d\Omega} = \frac{1}{4 E_1 E_4 \vert v_1-v_4 \vert}   \frac{\vert\bar{p}_2\vert}{16 \pi^2(E_1+E_4)}   \vert A(p'_1,p_2,p_3,p'_4) \vert^2
\end{equation}

Pulling everything together, we can now impose the particle content and limits used in section \ref{sec:lbqft}. Recall that states $ \lbrace \phi_3,\phi_4 \rbrace $ are scalars of mass $ m $, and $ \lbrace \phi_1,\phi_2 \rbrace $ are photons. Also suppose the scalar is stationary, but with $ m \gg E_1 $, corresponding to the centre of mass frame, meaning that $ E_3=E_4=m $. Next, we note that $ \vert v_1-v_4 \vert =1$ since the scalar is stationary and the photon propagates at the speed of light, which we all know is one. Lastly we recall that the change in energy of the photon is small, allowing us to write $ E_1 \approx E_2 = \vert \bar{p}_2\vert $. Plugging all of this into \eqref{eq:diffcross2} we get

\begin{equation}
\frac{d\sigma}{d\Omega} = \frac{1}{64 \pi^2 s_{14}}   \vert A(p'_1,p_2,p_3,p'_4) \vert^2 .
\end{equation}

This is a useful expression, however it is still not enough, since next we need to find some relation between this and the impact parameter. To this end consider very much the same set-up as before but with all particles classical. We have a cylindrical beam of equally distributed particles, with number density $ N_i $, incident on some other spherical target particle at the centre of the beam. The translational offset of the incident particles from the centre of the beam is measured by the impact parameter $ b $. 

So, some particles pass through a ring in the beam of width $ db $ situated at impact parameter $ b $. These particles proceed to scatter off the target particle into some solid angle region $ d\Omega = 2\pi \sin(\theta)d\theta $ situated at $ \theta $. From this, we can define the number of particles that pass through the ring as

\begin{equation}
dn_i = 2\pi b \,db N_i .
\end{equation}

Since all the particles incoming through $ db $ scatter into the solid angle $ d\Omega $, the number of particles scattered through this region is simply $ dn_s=dn_i $. The cross-section as defined before is the ratio of the number of scattered particles to the number of incident particles. Therefore the cross section is given by

\begin{equation}
d\sigma = \frac{dn_s}{N_i}  = 2\pi b\, db,
\end{equation}

which is simply the area the incoming particles pass through. This quantity per unit solid angle through which the scattered particles pass is then

\begin{equation}
\frac{d\sigma}{d\Omega} = \frac{b\, db}{ \sin\theta\, d\theta},
\end{equation}

and easily manipulated into the form necessary to calculate the scattering angle

\begin{equation}
\left( \frac{d\sigma}{d\Omega} \right) \sin\theta\, d\theta = b \,db.
\end{equation}

\section{A guide to the literature}

In appreciation of the fact that a substantial part of learning any new subject is to know {\it where} to find the information and, wanting to offer a more helpful answer than ``the arXiv", in this appendix, we collect some of the references that we have found useful in learning the subject ourselves. Our list is by no meant complete or even authoritative but we hope it will help draw the interested reader further into an exhilarating field.
\begin{itemize}
	\item \textit{Scattering Amplitudes in Gauge Theory and Gravity} by Elvang and Huang. By now, this has evolved into the standard reference on amplitudes. Based on a set of graduate lectures given at the University of Michigan, it contains plenty worked examples and useful exercises. While it focuses mainly on developments in gauge theory, it contains enough material on gravity to bring the reader up to speed on the frontiers of the field. It is available from Cambridge University Press, with a preprint available at \url{https://arxiv.org/abs/1308.1697}
	\item \textit{Tales of 1001 Gluons} by Stefan Weinzierl. A more modern introduction, slightly more up to date and covering some aspects that are omitted in Elvang \& Huang. In addition to and excellent section on perturbative gravity, it contains, for example, new material on the scattering equations and CHY representations. As a bonus, it also contains numerous exercises and solutions. making it an excellent guide for students. It can be found on the arXiv at \url{https://arxiv.org/abs/1610.05318}.
	\item \textit{A brief introduction to modern amplitude methods} notes by Lance Dixon. Another modern (although slightly older now) general introduction to the topic. Mostly focusses on gluon scattering/loops. \url{https://arxiv.org/abs/1310.5353}
	\item \textit{A First Course on Twistors, Integrability and Gluon Scattering Amplitudes} is based on a set of lectures given by
Martin Wolf at Cambridge University. It is decidedly for the more mathematically inclined readership but pedagogical and with some excellent references for anyone wanting to fall further down this particular rabbit hole. It can be found on the arXiv at \url{https://arxiv.org/abs/1001.3871}.
\item \textit{Lectures on twistor theory \& Scattering Amplitudes and Wilson Loops in Twistor
		Space} are also an excellent sets of pedagogical notes. They can both be found on the arXiv at \url{https://arxiv.org/abs/1104.2890} and \url{https://arxiv.org/abs/1712.02196}.
	\item \textit{Quantum Field Theory and the Standard Model} by Mathew Schwartz. This is primarily a text on quantum field theory geared toward the standard model of particle physics. As such, its focus is more on practical calculations rather than excessive formality. This makes it a very useful concise alternative introduction to QFT. Of particular interest to us is its introduction to spinor helicity methods which, although not as complete as the above texts, is certainly a useful suppliment.
	\item \textit{Quantum field theory} by Mark Srednikci. Another contemporary introduction to QFT. In addition to some slightly diffierent topics covered from other canonical texts, this one offers some useful examples of calculating cross sections and using the spinor helicity formulation.
	\item \textit{Quantum Field Theory in a Nutshell} by Anthony Zee. Now in its second edition, Zee's book has developed somewhat of a cult following among graduate students of many branches of physics. A very readable, if somewhat colloquial introduction to QFT, Zee develops the subject from the very basics and takes the reader all the way up to the frontiers of the field with many stops along the way to admire the scenery. While both editions are superb, it is the second edition in particular which contains a section on spinor helicity methods and a section on gravitational waves (both in 'Part N'). What it lacks in rigor, Zee's text more than makes up for in building intuition.
	\item \textit{Introduction to the Effective Field Theory Description of Gravity} by John Donoghue. Very good guide to perturbative gravity and effective field theories of gravity.\\ \url{https://arxiv.org/abs/gr-qc/9512024}
	\item \textit{Perturbative quantum gravity and its relation to gauge theory} by 't Hooft. Very concise introduction to Perturbative gravity by one of the masters in the field. 't Hooft's take on the subject, as always, brings with it a different perspective. An excellent reference source, it can be downloaded from \url{http://bit.ly/2m1rvsX}
	\item \textit{EPFL Lectures on General Relativity as a Quantum Field Theory}. A good introduction to various aspects of GR as viewed through the lens of quantum field theory.\\ \url{https://arxiv.org/abs/1702.00319}
\end{itemize}

\section{Glossary of terms}

{\bf Amplitude}: The probability amplitude for a certain interaction process of particles. In quantum field theory it is calculated by summing all the possible ways the interaction can be take place, which in Feynman diagram language are all the possible diagrams for the process allowed by the Feynman rules. An $n$-point amplitude is the amplitude of a process involving n physical particles.

{\bf BCFW Recursion Relation}: The BCFW recursion relations are a set of relations that allows one to construct multiple particle amplitudes from sub-amplitudes (amplitudes with a lower number of particles). That is, if one were to calculate an n-point amplitude one could write it as the summed product of $i$-point and $(n-i)$-point amplitudes with $i=2,...,n-2$.

{\bf Cross-Section}: The classical cross-section is the area transverse to the relative motion of two particles within which they need to be in order to interact. For hard spheres the cross section is the area of overlap of the objects in order to collide. In instances where the interaction is mediated by a potential the cross section is generally bigger than the actual particle.

{\bf Differential Cross-section}: The differential cross-section, $d \sigma/d \Omega$, is the ratio of particles scattered into a certain direction per unit time per unit solid angle divided by the the number of incident particles. This is equivalent to taking the normalized spin sum of a scattering process and can be related to the total cross section by integrating over all solid angles.

{\bf Feynman Rules}: The set of interaction rules for particles derived from the action of an appropriate theory. Includes vertices , propagators and external particle state contraction.

{\bf Helicity}: The helicity of a particle is the projection of the spin onto the linear momentum of the particle.

{\bf Impact parameter}: Commonly denoted by $b$, the classical impact parameter is defined as the perpendicular distance between a particle and the center of a potential field the particle scatters off of.

{\bf Little Group}: The little group is the set of transformations that leaves the momentum in a given direction of an on-shell particle invariant.

{\bf Mandelstam Variable}: These are gauge invariant quantities that are constructed from the four-momenta of the physical particles in a scattering process. More precisely they are the square of the sum of the momenta of all particles as a given vertex containing a propagator. And are generally denoted $s_{ijk...} = -(p_i^\mu + p_j^\mu + p_k^\mu + ...)^2$.

{\bf On/Off-shell}: Particles that are on-shell adhere to the equations of motion, hence satisfying a physical constraint that relates their momentum and energy. For a particle of mass $m$, this is the well-known relation in special relativity $p^{\mu}p_{\mu}= -m^2$. An off-shell particle does not satisfy this constraint, so that $p^{\mu}p_{\mu} \neq -m^2$.

{\bf Propagator}: In the language of quantum field theory the propagator is a virtual particle that transfers momentum between two particles that are interacting.

{\bf Scalar Particle}: Scalars are particles with spin 0. In this paper we use this particles to approximate stellar bodies such as the Sun.

{\bf Spin Sum}: The spin sum is the sum of the complex square of an amplitude with the sum ranging over all the possible spin-state configurations of the amplitude.

{\bf Spinor Particle}: Spinors are particles with spin $1/2$, commonly called fermions. Fermions can be represented as four component Dirac Spinors or as is used in the spinor-helicity formalism as two component Weyl spinors.

{\bf Vector Particle}: Vectors bosons, or gauge bosons, are particles with spin 1, like photons or gluons. They are represented mathematically using Lorentz vectors.

{\bf Vertex}: In Feynman diagrams the vertex is a point in which three or more particles interact in the diagram, and momentum is conserved at all vertices in the given diagram. The vertex is represented mathematically by the vertex expression that is derived from the the interaction Lagrangian. 

{\bf Virtual Particle}: A virtual particle is an off-shell particle used in Feynman diagrams to mediate the interaction between two or more interacting physical particles, i.e. to act as a way to transfer momentum/information between physical particles. In scattering amplitudes, it is always represented by a propagator.

\end{fmffile}
\newpage
\bibliographystyle{amps}
\bibliography{biblio}
\end{document}